\pdfoutput=1
\documentclass[iop,apj]{emulateapj}
\usepackage{amsmath}
\usepackage{color}
\usepackage{natbib}
\usepackage{graphicx}
\usepackage{multirow}

\newcommand{\kmsMpc}{km~s$^{-1}$~Mpc$^{-1}$}
\newcommand{\kms}{km~s$^{-1}$}
\newcommand{\Mpch}{$h^{-1}$~Mpc}
\newcommand{\hMsun}{$h^{-1}\;\mathrm{M}_\odot$}
\newcommand{\APz}{Alcock-Paczy\'nski }

\shorttitle{Precision cosmography with stacked voids}
\shortauthors{G. Lavaux \& B. D. Wandelt}

\begin{document}

\bibliographystyle{apj}

\title{Precision cosmography with stacked voids}

\author{Guilhem Lavaux\altaffilmark{1,2,5,6}}
\affil{Department of Physics, University of Illinois at Urbana-Champaign, 1002 W Green St, Urbana, IL, 61801, USA}

\author{Benjamin D. Wandelt\altaffilmark{3,1,4}}
\affil{UPMC Univ Paris 06, UMR 7095, Institut d'Astrophysique de Paris, 98 bis, boulevard Arago, 75014 Paris, France}

\altaffiltext{1}{Department of Physics, University of Illinois at Urbana-Champaign, 1002 W Green St, Urbana, IL, 61801, USA}
\altaffiltext{2}{Department of Physics and Astronomy, The Johns Hopkins University, 3701 San Martin Drive, Baltimore, MD 21218, USA}
\altaffiltext{3}{CNRS, UMR 7095, Institut d'Astrophysique de Paris, 98 bis, boulevard Arago, 75014 Paris, France}
\altaffiltext{4}{Department of Astronomy, 1002 N Gregory Street, University of Illinois at Urbana-Champaign, Urbana, IL 61801, USA}
\altaffiltext{5}{Department of Physics and Astronomy, University of Waterloo, 200 University Avenue West, Waterloo, Ontario, Canada, N2L 3G1}
\altaffiltext{6}{Perimeter Institute for Theoretical Physics, 31 Caroline Street North, Waterloo, Ontario, N2L 2Y5}

\begin{abstract}
We present a purely geometrical method for probing the expansion history of the Universe from the observation of the shape of stacked voids  in spectroscopic redshift surveys. Our method is an \APz test based on the average sphericity of voids posited on the local isotropy of the Universe. It works by comparing the temporal extent of cosmic voids along the line of sight with their angular, spatial extent. We describe the algorithm that we use to detect and stack voids in redshift shells on the light cone and test it on mock light cones produced from $N$-body simulations. We establish a robust statistical model for estimating the average stretching of voids  in redshift space and quantify the contamination by peculiar velocities. Finally, assuming that the void statistics that we derive from $N$-body simulations is preserved when considering galaxy surveys, we assess the capability of this approach to constrain dark energy parameters. We report this assessment in terms of the figure of merit (FoM) of the dark energy task force and in particular of the proposed EUCLID mission which is particularly suited for this technique since it is a  spectroscopic survey. The FoM due to stacked voids from the EUCLID wide survey may double that of all other dark energy probes derived from EUCLID data alone (combined with Planck priors).  In particular, voids seem to outperform Baryon Acoustic Oscillations by an order of magnitude.  This result is consistent with simple estimates based on mode-counting. The \APz  test  based on stacked voids may be  a significant addition to the portfolio of major dark energy probes and its potentialities must be studied in detail.
\end{abstract}

\section{Introduction}

The physical nature of Dark Energy, detected through  supernovae luminosity distance measurements \citep{UNION08},  Baryonic Acoustic Oscillations \citep[BAO, ][]{BAO10} and the Cosmic Microwave Background \citep{WMAP7_K11},  still evades us. The BOSS survey \citep{BOSS} has been designed to assess whether the equation of state of Dark Energy is indeed constant and equal to minus one. However, observations based on baryonic acoustic oscillations are limited by the minimal volume required to estimate the scale of these oscillations, typically $\sim$100\Mpch{}. With the advent of large galaxy spectroscopic redshift survey, such as the Sloan Digital Sky Survey \citep{SDSS_DR7}, we now have access to a three-dimensional representation of the large-scale structure on our light-cone on vastly different scales. 

The well known \APz test \citep{AP79} can be applied to any structure for which we know the physical size or, more weakly, the ratio of its extent  along the line of sight and its angular size. In particular, if we had a population of standard spheres scattered throughout cosmic history we could  measure the cosmological expansion directly. Absent such a population, the next best thing is a population of objects whose average shape is spherical. 

Cosmic voids are such a population and hence promising candidates for probing the expansion geometry of the Universe. Even though individual void shapes may be complicated, \textit{the average void is spherical} in an isotropic and homogeneous universe. Detecting all voids observed in a galaxy survey and stacking voids of similar sizes and redshifts projects out the details of individual void shapes.   Since the average shape of voids is known to be spherical, the observed, stretched shape in redshift space is a direct function of the local Hubble expansion of the Universe and  the angular diameter distance at the redshift of the void and hence a sensitive function of the cosmological parameters, in particular those parameterizing the dark energy equation of state.

Voids are spatially localized coherent structures with sizes between  $\sim$5-20 Megaparsecs.  If voids can be used to construct an \APz test, they will unlock a much larger number of modes for precision cosmology than are accessible by the BAO technique, including modes that are in the mildly non-linear regime.  

Crucially, by focusing on void regions we take advantage of the much more easily modeled phase-space structure in low-density regions compared to high-density regions. 

The promise of an AP test based on voids was first noticed by \cite{Ryden95} who proposed using the apparent stretching of void shapes in redshift space coordinates to estimate the local geometry of the expansion. Doing so properly requires selecting voids that have the same overall size and density. This work has then yielded a number of other studies on voids and baryonic acoustic oscillations \citep{BPH96,RM96,SRM01}, though the complicated shapes of individual voids have made it difficult to extract cosmological information at high signal-to-noise.

In a similar spirit, \cite{JL02}  proposed a test which uses cosmic clocks as tracers of cosmic time to which we can compare the measured galaxy redshifts. The \APz test can be thought of as a differential version of this approach---one could say comparing the radial, temporal extent of voids with their angular, spatial extent amounts to using them as ``cosmic stopwatches.'' Further, it does not rely on spectral modeling to extract galaxy ages.

\cite{PK10} proposed using genus statistics to measure expansion. Their claim was that the genus is insensitive to redshift space distortions while peculiar velocities will mildly  affect the void technique \citep{BPH96}. However, the genus is not a spatially local quantity, which may be prone to a number of observational problems like non-trivial edge effects and inhomogeneous incompleteness corrections. It is also model dependent.

Our method requires spectroscopic survey since redshift errors in photometric redshift catalogs wash out the line-of-sight information on the scale of all but the most extreme voids even with $\sim 0.7\%$ precision \citep[e.g.][]{I09}. An order of magnitude improvement in redshift precision would likely be required to directly observe any non-linear three-dimensional structures (\textit{cf.} \cite{JW11} for a possible approach).

\begin{figure*}
  \begin{center}
 	\includegraphics[width=.8\hsize]{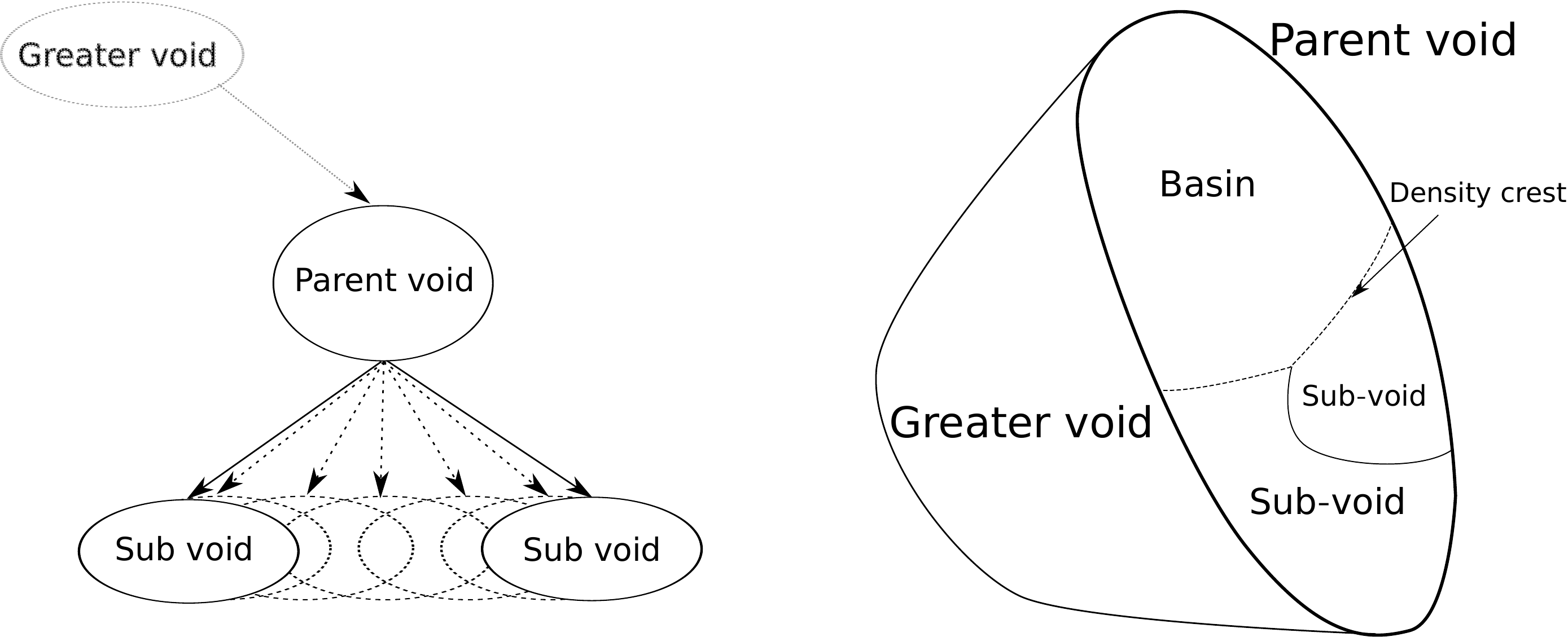}
  \end{center}
	\caption{\textit{Storing voids in a hierarchical tree.} The spatially adjacent sub-voids are assembled in a single void, the ``parent void'', which contain at least one additional basin. The identity of the parent void is inherited by the ``sub-void'' with the smallest core density. The parent void is itself part of a ``greater void''. The tree is ordered in scale.}
\end{figure*}

Our paper is organized as follows. In Section~\ref{sec:cosmo_with_voids}, we show how voids may give us a direct probe of cosmological parameters through shape stretching. In Section~\ref{sec:finding_voids}, we explain the method  we use for finding and stacking  voids on  the light-cone  and inferring the local expansion from the shape of stacked voids. In Section~\ref{sec:test_nbody}, we test our method on $N$-body simulations. We derive the profiles of stacked voids in simulation, the sensitivity to contamination of the redshifts by peculiar velocities and the number density of voids found in the simulation. In Section~\ref{sec:hubble}, we derive the Hubble-diagram of expansion in the simulation from our mock-observation of voids. We do a  Fisher-matrix analysis of the measurement of Dark Energy physical parameters using the expansion rate derived from voids. We apply this formalism to the survey specifications of the main galaxy sample of the Sloan Digital Sky Survey \citep[SDSS][]{SDSS_DR7}, the Baryonic acoustic Oscillation Sky Survey \citep[BOSS][]{BOSS} and the EUCLID survey \citep{EUCLID2}. In Section~\ref{sec:conclusion}, we conclude.

\section{Cosmology with voids}
\label{sec:cosmo_with_voids}

In this Section, we recall the basic equation at the base of the Alcock-Pasczy\'nski  test \citep{AP79} applied on voids. This test comes comes from the relation between the comoving angular distance $D_A$ and the redshift $z$ of an event in a Friedmann-Lema\^itre-Robertson-Walker (FLRW) cosmology:
\begin{equation}
  D_A(z) = \frac{c}{H_0} f_k\left(\frac{H_0}{c}\chi(z)\right), \label{eq:cosmo_redshift}
\end{equation}
with
\begin{equation}
  f_k(x) = \left\{
     \begin{array}{ll}
       \frac{1}{\sqrt{|k|}} \sinh(\sqrt{|k|}x) & \mathrm{if }\;k < 0 \\
       x & \mathrm{if }\; k = 0\\
         \frac{1}{\sqrt{|k|}} \sin(\sqrt{|k|} x) & \mathrm{if }\; k > 0
     \end{array}\right.,
\end{equation}
the redshift/comoving distance relation
\begin{equation}
  \chi(z) = \frac{c}{H_0} \int_0^z \frac{\mathrm{d}\tilde{z}}{E(\tilde{z})}, \label{eq:chi_def}
\end{equation}
where $E(z) = H(z)/H_0$, and
\begin{equation}
  k = \left(\frac{H_0}{c}\right)^2 (\Omega_\text{m}+\Omega_\Lambda-1),
\end{equation}
with $\Omega_\text{m}$, the mean matter density, and $\Omega_\Lambda$ the Dark Energy density, both normalized to the present critical density.
For sufficiently small curvatures $k$ , it is possible to invert the relation~\eqref{eq:cosmo_redshift} to derive the redshift from the comoving distance $r$
\begin{equation}
  z = D_A^{-1}\left(\frac{r H_0}{c}\right). \label{eq:inv_cosmo_redshift}
\end{equation}
If we look at a cosmic object, {\it e.g.} a galaxy, or a cosmic void or the BAO, at redshift $z$, it has an extent $\delta z$ in the redshift direction and $\delta r$ in the angular direction defined as
\begin{equation}
	\delta r \equiv D_A(z) \delta \theta.
\end{equation}
Additionally $\delta z$ corresponds to a comoving distance along the line-of-sight given by the differentiation of Eq.~\eqref{eq:cosmo_redshift}:
\begin{equation}
  \delta l = \frac{\mathrm{d}D_A}{\mathrm{d}z} \delta z = \frac{c \delta z}{H_0 E(z)} f'_k(\chi(z)),
\end{equation}
with $f'_k$ the first derivative of $f_k$. We have indicated in the introduction that we assume the Universe is locally isotropic. Consequently, the large-scale structures must not have a preferred direction in average. If we consider a void consisting in an infinite average number of stacked voids of a specific volume, this ``stacked void'' should have the same extent in all directions. We can thus assume that $\delta l = \delta r$. This equality yields
\begin{equation}
  \frac{\delta r}{\delta z} = \frac{c}{H_0 E(z)} f'_k(\chi(z)), \label{eq:local_hubble_comoving}
\end{equation}
which in terms of the projected separation $\delta d = c z \delta \theta / H_0$ gives
\begin{equation}
  \frac{\delta z}{\delta d} = \left(\frac{H_0}{c}\right)^2 \frac{D_A(z) E(z)}{z f'_k(\chi(z))} = \frac{H_0}{c} e_\mathrm{v}(z). \label{eq:hubble}
\end{equation}
We propose observing this quantity through measuring the shape of stacked voids in redshift space as a function of redshift. Note that this observable depends directly on $E(z)$ rather than through an integral, as is the case for methods based on angular diameter distance, probed by observing the angular scale of the BAO peak as a function of redshift, or the luminosity distance which is probed by supernova surveys. We expect this to enhance the sensitivity of our method to the physical properties of dark energy, which appear directly in E(z).
This relation is not strictly an image of the Hubble constant at different $z$ as it is modified by $D_A(z)/(z f'_k(\chi(z)))$, which is close to one at low redshift. However, it is a good proxy for it and it is possible to obtain the equivalent of an Hubble diagram for voids. Eq.~\eqref{eq:hubble} was already derived by \cite{Ryden95} for universes with no curvature. The measurement of the isotropy would be a clean way to measure finely the cosmic expansion.

\section{Finding and stacking voids}
\label{sec:finding_voids}

In this Section, we present the algorithm that we developed to locate voids and stack voids on an expanding metric. In the following, we will use the following convention. The effective radius of a void corresponds to the radius of the sphere of equivalent volume. So, if $V$ is the volume of the void,
\begin{equation}
	r_\mathrm{eff} = \left(\frac{3}{4\pi} V\right)^{1/3}.
\end{equation}

First, in Section~\ref{sec:coords}, we define the coordinate system that we use in this work. In Section~\ref{sec:void_tree}, we give the details of the algorithm to find and choose the voids that are of interest for the stacking procedure. In Section~\ref{sec:void_stacking}, we describe the stacking procedure.

\subsection{The coordinate system} 
\label{sec:coords}

The fundamental point on which we base our method is the capability to find and stack void structures, even if they are strongly distorted. For simplicity, we adopt the following definition, which a posteriori we will show is robust to distortions in the coordinate system. We will use the infinite remote observer approximation for the redshift coordinate and the planar approximation for the angular coordinate. We consider density tracers, e.g. galaxies, in a hybrid coordinate system $(x,y,z)$.  For a tracer $t$, {\it e.g.} a galaxy, at the sky position $(\theta,\phi)$, $\theta$ being the sky latitude and $\phi$ the sky longitudes, and located at redshift $Z$, we define:
\begin{align}
	x&= \frac{c Z}{H_0} \cos(\phi) \cos(\theta) \\
	y&= \frac{c Z}{H_0} \sin(\phi) \cos(\theta) \\
	z&= \frac{c Z}{H_0} \sin(\theta)
\end{align}
In the infinite remote observer approximation $\theta\sim \pi/2$ and thus $z \sim \frac{c Z}{H_0}$, but other directions are adequate provided that the extent on the sky is small.

\subsection{Organizing voids in tree}
\label{sec:void_tree}

\begin{figure}
  \includegraphics[width=\hsize]{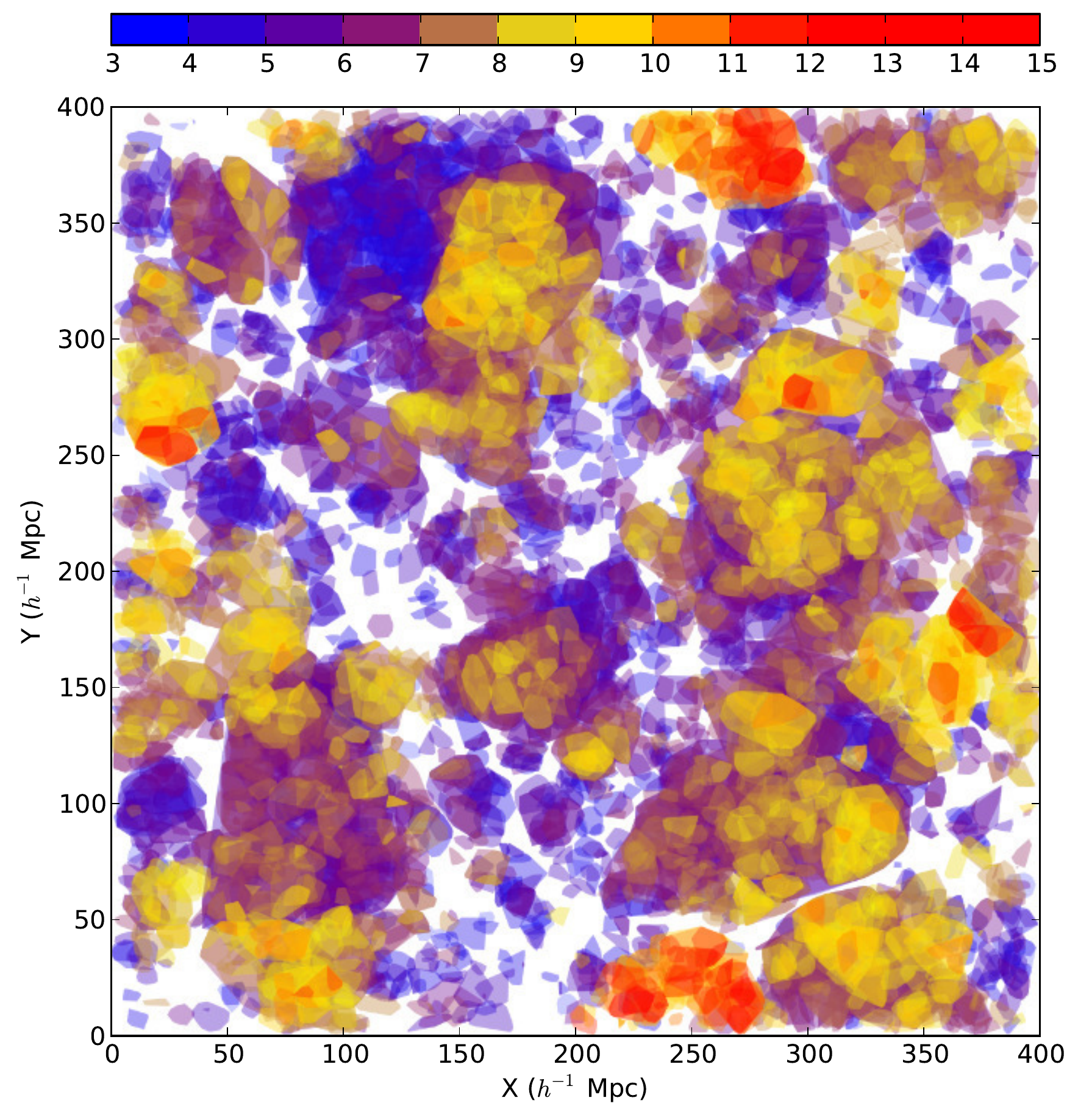}
  \caption{\label{fig:tree_figure} \textit{Void tree from an $N$-body simulation.} We show the convex hulls of particles within a 40\Mpch{} deep slice, and belonging to a void. A void is visualized  if it has either a mean relative density less than $-0.4$, for clarity of the representation, and an effective radius within $[8;50]$\Mpch{}, or is within a subtree of an drawn void. The color encodes the depth of the void in the tree as indicated by the top color bar.   }
\end{figure}

From the volume sampled by the tracers $\{ t \}$, we extract a box of side $L$ in the $(x,y,z)$ coordinate. We now use the {\sc Zobov} \citep{N08} algorithm to compute and locate local minima  in the density of tracers, assuming they are a sample of the underlying matter density field. {\sc Zobov} finds local density minima on density field sampled by particles and their associate catchment basins, in the language of the watershed transform \citep[e.g.][]{WVF}. It does so using a Voronoi tessellation, derived from the Delaunay tessellation applied on a set of tracers. In addition, basins are assembled in voids, starting from the lowest density basin, such that:
\begin{itemize}
	\item[-] each basin is a void
	\item[-] two basins are assembled in one void if they share a common boundary, and that the density on this boundary is the lowest for each of the void. Basins are always assigned to the voids which have the lowest core density.
\end{itemize}
The whole volume sampled by mass tracers is thus partitioned into a set of basins. Each basin corresponds to a void which itself is a collection of basins. Thus, the voids naturally acquire  a tree-like structure for which voids have both a single ``parent'' and possibly many ``children''. We define an order on the tree such that a void is the immediate parent of another one if it shares the same zones as the child void and at least one more. We note that this corresponds to a different tree than defined by \cite{AWAPS10}. Also, contrary to many galaxy based void finder, this method does not assume a hard-coded void shape \citep[see ][ for a review of void finders]{C08}. It does not  rely on a smoothed density field but only on the topology of the tracers, such as particles in $N$-body simulations or galaxies in observations. This is not the sole technique. More recently the complete topology of the cosmic web which can be derived through a set of tracers has been formally studied by \cite{S11} in the context of the persistence. The tree that we have developed helps separating overlapping voids such that the same volume is not used multiple times in a statistical analysis.

 In our approach, there may not be a natural single root for this tree because we do not have periodic boundary for the box side $L$. So we introduce an artificial root node for which all child nodes correspond to the parent-less voids. Additionally, we compute the mean density of each void in the tree, which is a non-monotonous function of the depth in the tree.

\subsection{Void stacking}
\label{sec:void_stacking}

\begin{figure*}
  \begin{center}
  \includegraphics[width=.45\hsize]{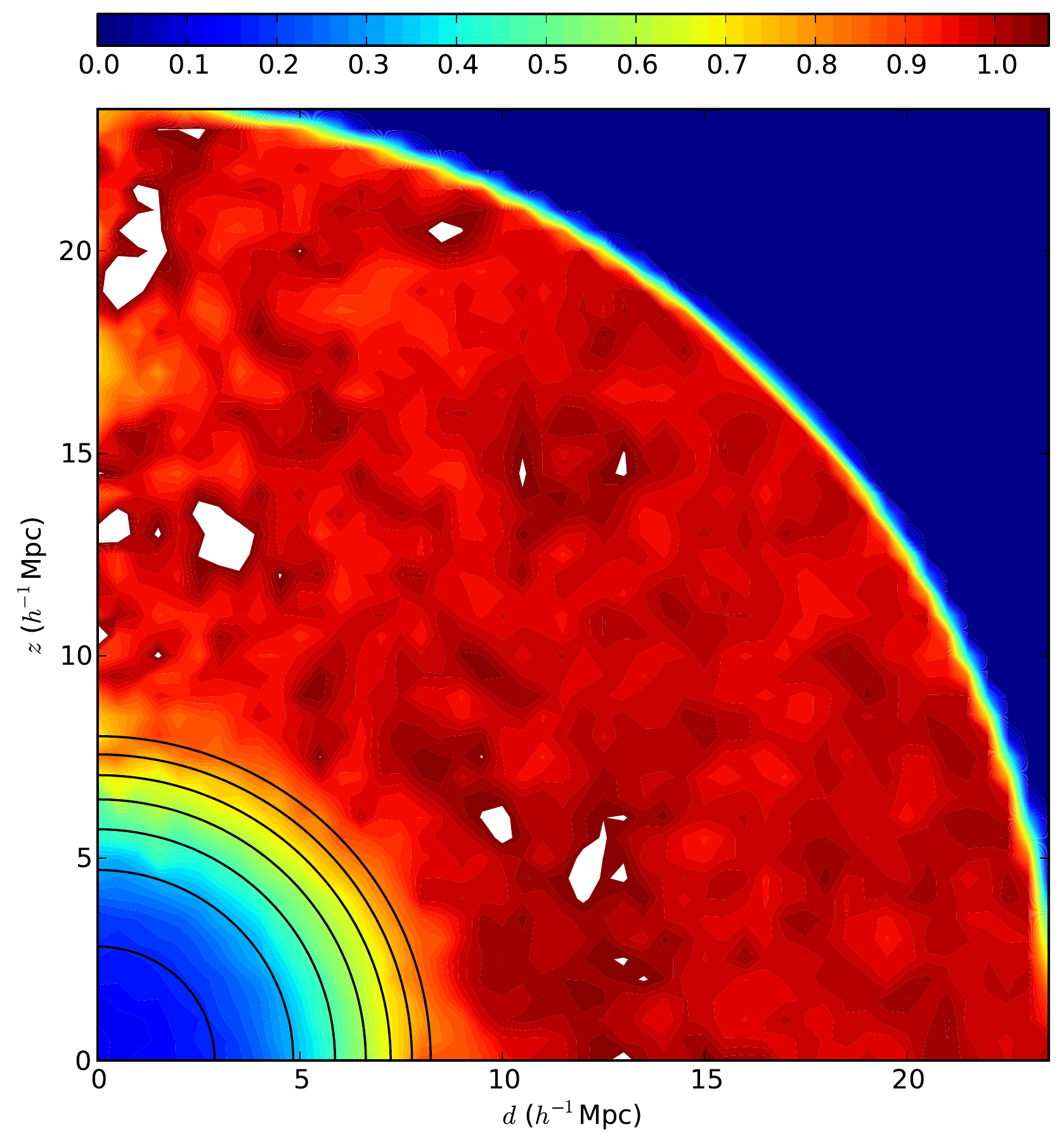}
  \includegraphics[width=.45\hsize]{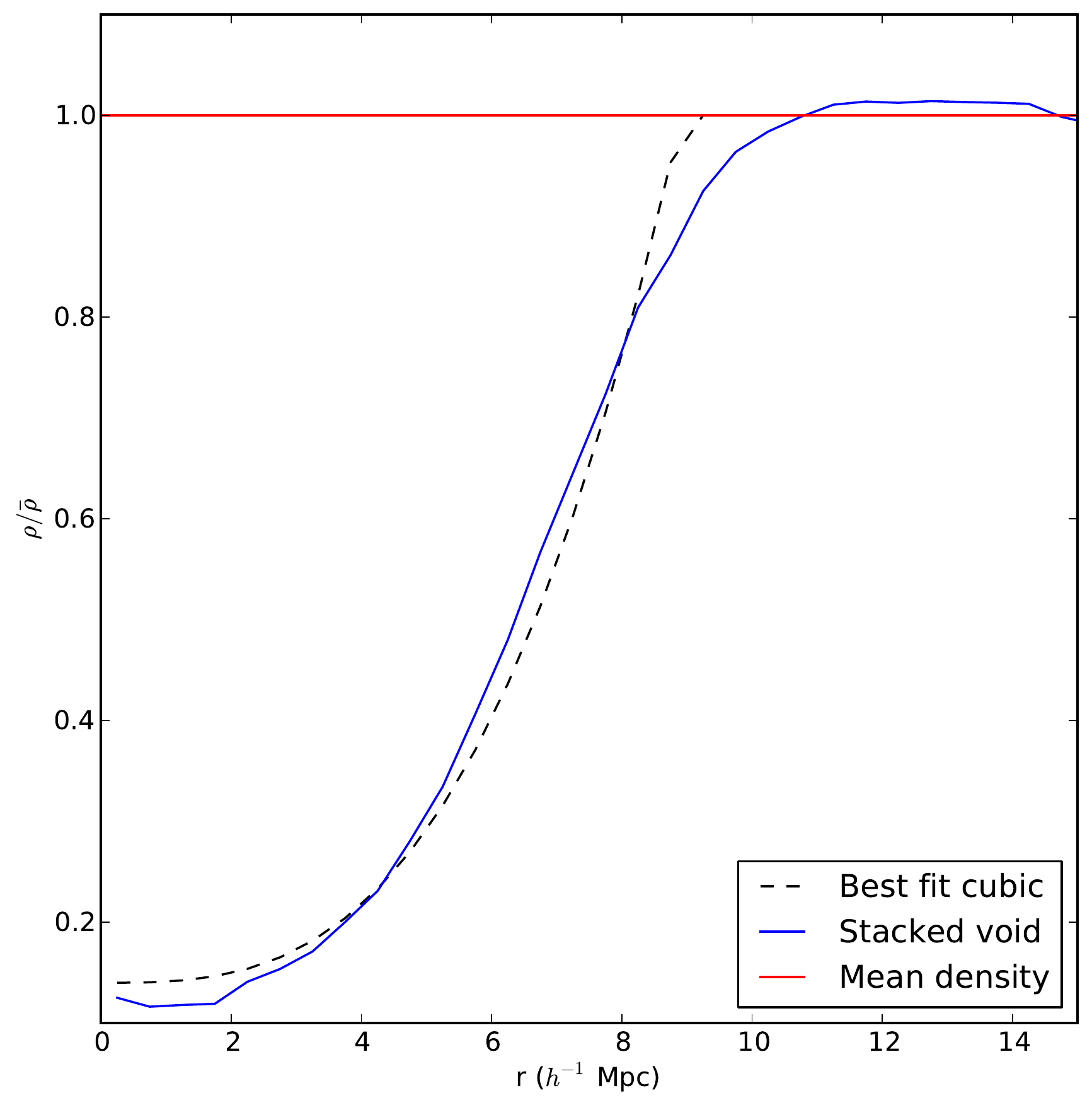}
  \end{center}
  \caption{\label{fig:stacked_void} {\it Void stacking in the final, $z=0$ state of the simulation.} We show  the result of the stacking procedure given in Section~\ref{sec:void_stacking} when it is applied to all voids of size 8 Mpc in the full volume of the simulation without introducing cosmological and peculiar velocities distortions. Left panel: a filled contour plot of the density $n(d,z)$. Contours above the density $1.2$ are shown in light gray. We fitted the simple ellipsoidal model of Section~\ref{sec:ellipsoidal} on the binned density to estimate the ratio between the axis along the $XY$ direction (x-axis) and the axis along the $Z$ direction (y-axis). Right panel: the three-dimensional average density profile, in shells, of the stacked void (solid line) and the fit to a cubic density profile (dashed line). The horizontal solid red line is the mean density.}
\end{figure*}

\begin{figure*}
   \begin{center}
	\includegraphics[width=\hsize]{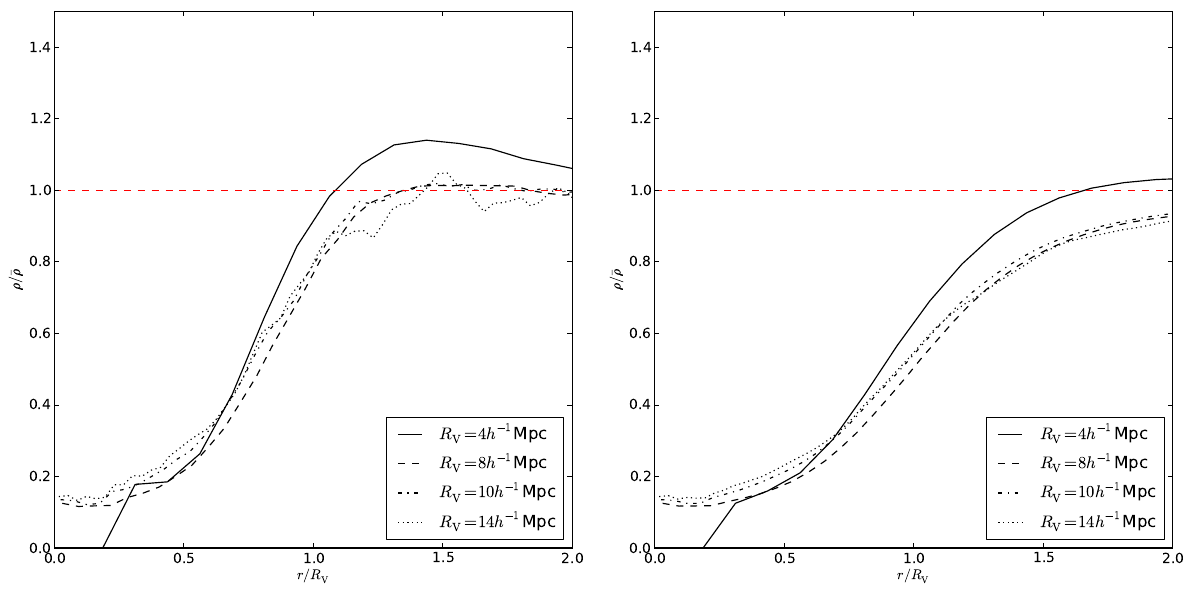}
   \end{center}
	\caption{\label{fig:all_profiles} {\it Stacked void profiles at $z=0$} -- We show the profiles for voids selected with different effective radii. Left panel: mass density profiles in thin shells. Right panel: average mass density profile in the sphere of the given radius.}
\end{figure*}

Voids have complicated shapes mainly produced by the gravitational shear field \citep{PL07,LW10}. It is therefore difficult to use them individually  as a probe of the effect of cosmological expansion. However, using the assumption of the isotropy of density fluctuations, we expect that the average void shape is spherical in physical coordinates. To achieve a correct stacking, at least three important features are required:
\begin{enumerate}
\item for a given stack, the algorithm must only select voids within a narrow volume range. This condition will cause a specific average void shape of the given size to emerge from the stacking. 
\item each void of a stack must be unique and must not overlap with another void. Enforcing this condition allows the removal of spurious correlation in the void shapes, which could systematically affect the result. 
\item each void in a stack must be centered precisely on its \emph{average} lowest density. This is necessary to minimize the effect of halos, which are expected on the boundaries of voids. These halos may bias the position of the center reducing the amount of signal available from the average void shape. 
\end{enumerate}
The first  requirement is satisfied by the {\sc Zobov} void finder.  {\sc Zobov} provides a volume for the void which is exact from the point of view of topology. The second point is achieved by storing the voids in the tree structure mentioned at Section~\ref{sec:void_tree}.

The center of the void could be derived directly from the Voronoi tessellation. However we expect this inferred center to be unstable with respect to shot noise in observations, or the Lagrangian grid in $N$-body simulations. These would cause  spurious effects in the stacking procedure. We opt for computing the mean lowest density position using the volume-weighed barycenter of the tracers attached to a single void of the stack. The Voronoi tessellation gives a volume $V_i$ surrounding each tracer $i$. We compute the average position $\mathbf{x}_\mathcal{V}$ of the center of the void $\mathcal{V}$ by computing 
\begin{equation}
	\mathbf{x}_\mathcal{V} = \frac{1}{\displaystyle \sum_i V_i} \sum_i \mathbf{x}_i V_i\label{eq:volbary},
\end{equation}
where $i$ runs over the tracer, {\it e.g.} galaxies, in the void $\mathcal{V}$, and $\mathbf{x}_i$ its position in the coordinate system given in Section~\ref{sec:coords}.

Finally, some voids of the stack have large clusters near their estimated center. In the limit of an infinite number of voids, we expect these cases to be vanishingly small. However, we do not have an infinite number of voids in reality. We have decided to avoid these cases by enforcing that the core density of the void should be really empty. We define the core density as the mean matter density within a sphere of fiducial radius $r_\mathrm{eff}/4$. This comes at the cost of a smaller number of available voids in the stack. A posteriori, in Section~\ref{sec:stack_comoving}, we see that our adopted fiducial radius is smaller than the actual size of the ``core'' of the stacked void. 

To summarize, we proceed as follows:
\begin{enumerate}
\item We put the tracers in the coordinate system of Section~\ref{sec:coords}.
\item We extract a parallelepiped volume of side $L\times L \times L_z$ which  resides entirely within the region spanned by the tracers. 
\item We normalize the coordinates to 1 in each direction.
\item We apply the {\sc Zobov} void finder algorithm.
\item We store the found voids in a tree according to Section~\ref{sec:void_tree}.
\item We walk the tree, starting from the root and stop whenever the effective radii, $r_\mathrm{eff}$, is within a given range,  between $R_\mathrm{min}$ and $R_\mathrm{max}$. $r_\mathrm{eff}$ is obtained by computing the radius of the sphere which has the same volume as the void.
\item We compute the position of the particles according to the volume-weighed barycenter of Eq.~\eqref{eq:volbary} for each void.
\item We compute the density of the void within a sphere of radius $r_\mathrm{eff}/4$. We only accept voids that have a core density less than 20\% of the mean matter density of the universe.
\item Around each volume-weighed barycenter, we extract a spherical volume, in the coordinates of Section~\ref{sec:coords}, of radius $R_\text{cut}=3\times R_{max}$. The center of the extracted volume is put at the origin. We do this for all selected voids.
\end{enumerate}
The resulting particle distribution gives what we call a stacked void. We transform coordinates from $(x,y,z)$ to $(d=\sqrt{x^2+y^2},|z|)$. We bin this distribution and divide the resulting density by the number of stacked voids, the bin width and $d$, the Jacobian of the transformation. This procedure yields the density $n(d,z)$ of tracers per unit volume per void. An example of such a distribution is given in Figure~\ref{fig:stacked_void}. 

\section{Tests on $N$-body simulation}
\label{sec:test_nbody}

We have run a series of three pure dark matter $N$-body simulations with different realizations of the initial conditions but for the same cosmology. The dark matter particles have different sampling properties than the galaxies. In particular, galaxies are biased tracers of the matter density field. As we are relying on topological properties of the large-scale structures, the result should be the same for the two. We discuss in Section~\ref{sec:conclusion} the limits of our approach. The volume of each simulation is given by a cube of side $L=500$\Mpch{}. Each simulation has $N=512^3$ particles. We have adopted a $\Lambda$CDM-WMAP7 cosmology with the 
following parameters: $\Omega_\mathrm{b} h^2 = 0.02258$, $\Omega_\mathrm{c} h^2=0.1108$, $H = 71$~\kmsMpc, $w=-1$, $n_\mathrm{S}=1$, $A_\mathrm{S}=2.34\times 10^{-9}$. This corresponds to 
$\Omega_\mathrm{b} = 0.045$, $\Omega_\mathrm{M}=0.264$, $\sigma_8=0.84$. Each particle has a mass $m_\mathrm{p}=2.05\,10^{11}$\hMsun.  The transfer function for density fluctuations for this cosmology is computed using {\sc CAMB} \citep{CAMB}. The initial conditions are generated using {\sc ICgen},\footnote{Available from http://www.iap.fr/users/lavaux/.} which uses the transfer function to generate a density field from the primordial power spectrum.

\subsection{Stacking voids in pure comoving coordinates}
\label{sec:stack_comoving}

In this Section, we consider the ideal case of voids stacked in comoving coordinates,\textit{i.e.} we purely consider the distribution of dark matter particles as given by the $N$-body simulation. We consider voids for which $r_\mathrm{eff}$ is between $R_\text{min}=8$\Mpch{} and $R_\text{max}=9$\Mpch{}. 

We give the result of the void stacking procedure for one of the $N$-body sample realizations in Figure~\ref{fig:stacked_void}. In the left panel, we show the  density profile of the stacked void, where on the horizontal axis corresponds to $d=\sqrt{x^2+y^2}$, with $x$ and $y$ the first and second coordinate of the particle in the stacking, and the vertical axis to $|z|$, the third coordinate of the particle. The solid black contour corresponds to the result given by the likelihood analysis described in Section~\ref{sec:ellipsoidal}. The filled color contours have been chosen equi-spaced from $\rho/\bar{\rho}=0$ to $\rho/\bar{\rho}=1.1$. The solid line in the right panel shows the three-dimensional density profile, in thin shells, of the stacked void. 

The one-dimensional density profile shown in the right panel of Figure~\ref{fig:stacked_void} is similar to the one shown in Figure 2 of \cite{CPVL06}, though we use pure dark-matter simulation in place of mock/real galaxy samples. The matter shell around the void is clearly visible for radii greater than the chosen $R_\text{min}$. 

The inspection of both panels shows a number of interesting features. It is clear that the voids stack coherently and form a region of lower density for $\sqrt{d^2+z^2} < 8$\Mpch{}. The mass density inside the stacked void, in both panels, is featureless. Outside the expected boundary of the stacked void, at $\sqrt{d^2+z^2} > 8$\Mpch{}, the density profile continues upward to $\rho\simeq 1.1 \bar{\rho}$ and then falls back to homogeneity. This shell is clearly seen in the one-dimensional averaging of the density profile shown in the right panel of the same Figure. Our density profile has similar features to the ones presented by \cite{BHTV03}, \cite{PCL05} and \cite{CPVL06}. 
In the two-dimensional mass density diagram of the left panel of Figure~\ref{fig:stacked_void}, it is clear that this is a real spherical shell, and that it harbors clumps highlighted by the white regions at $\rho/\bar{\rho} > 1.3$.

Our profiles are not as steep as in observations and some mock galaxy catalogs \citep[e.g.]{BHTV03, PCL05, HV04}. This may be due to our use of pure dark-matter simulation in place of mock galaxy samples. The results of \cite{BHTV03} (Figure 11) also indicates that observing fainter galaxies tends to smooth the density profiles, as expected.  Even though for the purpose of applying to observations \citep[such as in][]{CPVL06} it may be better to use their form, we have empirically found that a cubic function  fits the simulated density profiles adequately for the purposes of this analysis. We use the following form
\begin{equation}
	\frac{\rho(r)}{\bar{\rho}} = A_0 + A_3 \left(\frac{r}{R_\mathrm{V}}\right)^3 \label{eq:void_profile}
\end{equation}
with $R_\mathrm{V}$ the radius of the stacked void, $A_i$ the parameters against which the fit is computed. As we have voids with $8$\Mpch$ \leq r_\mathrm{eff} \leq 9$\Mpch{}, we have set $R_\mathrm{V}=8$\Mpch{}. The best fit gives $A_0=0.13\pm 0.01$, $A_3=0.70 \pm 0.03$ for voids of 8\Mpch{}.  

First, we note that the value of $A_0$ attributes a significantly non-zero density to the center of the void. This may be due to a resolution effect as putting one particle within $(1 h^{-1} \mathrm{Mpc})^3$ yields a density fluctuation of $\delta=-0.2$ for the resolution of our simulation. On the other hand, this value is the same for large voids where resolution effects should be milder. This  was already noted by \cite{CSDGY05}. 

\begin{figure*}
   \begin{center}
	\includegraphics[width=\hsize]{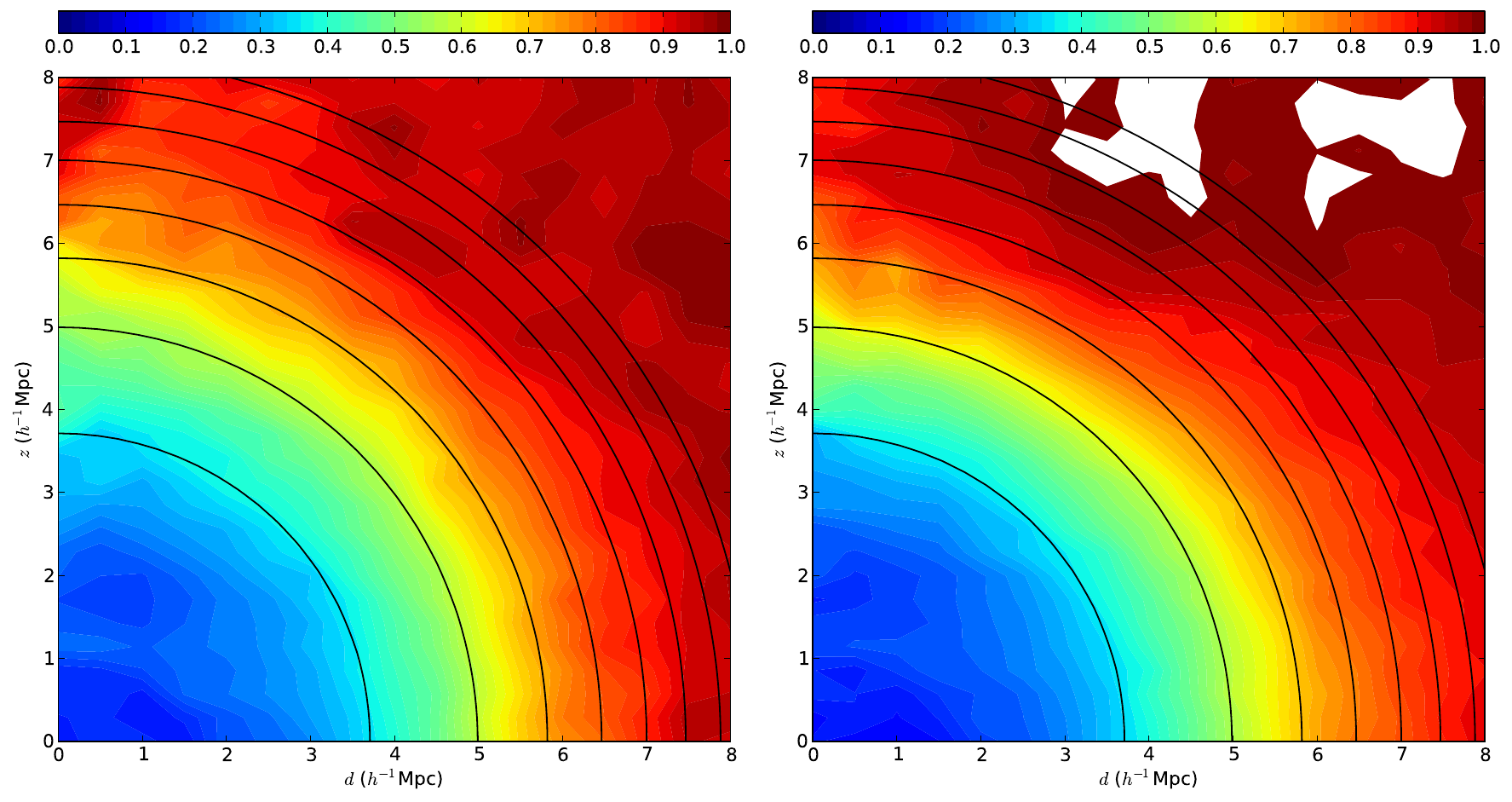}
   \end{center}
	\caption{\label{fig:redshift_vs_pure_expansion} 
		{\it Impact of peculiar velocities.} Density in  stacked voids in redshift coordinates, in the snapshot at $z=1$, for a 400\Mpch{}$\times$400\Mpch{} slice with a thickness of 10,000~\kms{}. Left panel: expansion only, without the peculiar velocities in the redshift positions of particles. Right panel: we have introduced the distortion due to peculiar velocities. The black contours are the same in the two panels and highlight the fitted density profile to the left, expansion-only case.	}
\end{figure*}

Second, we note that $(1-A_0)/A_3$ is not strictly equal to one. This can probably be explained by the size of the interval of accepted effective radii. We expect that large bin sizes, such as in \cite{CSDGY05}, may significantly change  the profile of the stacked void, notably at $r\sim R_V$ owing to the non-commutativity of the operation of re-scaling and averaging. This problem  also affects all the quantities that we may derive from this profile, like peculiar velocities. The result of the fit (dashed line) and the actual void profile (solid line) is shown in the right panel of Figure~\ref{fig:stacked_void}. Visual inspection shows a good agreement between those two profiles within the bounds of the voids. We will use this profile for the remainder of this work. For our method the choice of the profile is only important insofar as it does not bias (or remove signal to noise from)  the shape measurement.

In Figure~\ref{fig:all_profiles}, we show the density profiles, both in shells and cumulative, with radii normalized to the sought effective radii of the voids in each stack. We have considered voids with $r_\mathrm{eff}$ between $4-5$\Mpch{}, $8-9$\Mpch{}, $10-11$\Mpch{} and $14-15$\Mpch{}. We clearly see that all voids with $r_\mathrm{eff}$ greater than 8\Mpch{} have nearly exactly the same density profile, even at radii larger than $R_\mathrm{V}$. While the universality of void profiles was previously noted by \cite{CSDGY05}, they did not show that this universality  also extends outside the void, in the shell region. Our density profiles compares well also with that in the seminal paper by \cite{WK93}. Note that their void size definition is radically different from ours. Finally, from the left panel of Figure~\ref{fig:all_profiles}, we note the presence of a structure looking like a ``core'', which does not extend above half the effective radii of the individual voids. This justify a posteriori our choice for the mean density within $r_\mathrm{eff}/4$ to be sufficiently low to accept a void in the stack. 

\subsection{Stacking with redshift distortion}
\label{sec:stack_all}

In this section we consider the possibility of applying the algorithm for void detection of Section~\ref{sec:finding_voids}.
We have selected five snapshots for each of the three simulations. They correspond to the simulated universes at $z_\text{b}=0$, $z_\text{b}=0.25$, $z_\text{b}=0.54$, $z_\text{b}=0.81$ and $z_\text{b}=1.0$. We have considered that the first two dimensions of the positions of the particles in the snapshots correspond to the angular coordinates and the last, $z_C$, to the distance along the line of sight, relative to the face with $z_C=0$. As our simulated universes are flat, we have converted the comoving positions $z_C$ in redshift positions $z$ using
\begin{equation}
  z = \chi^{-1}\left(\frac{H_0}{c} \left(z_C + \chi(z_\text{b})\right)\right) + \frac{v_{z}}{c},
\end{equation}
with $\chi$ as defined in Eq.~\eqref{eq:chi_def}, $v_z$ the peculiar velocity of the particles in the $z$ direction, $c$ the speed of light. We have not included the additional $D_A(z)/z$ term present in $e_v(z)$ (Eq.~\ref{eq:hubble}) which adds an additional small effect at low redshift. In each case, we have extracted a box of 400\Mpch{}$\times$400\Mpch{}$\times$40,000~\kms{} from the distribution of particles in redshift coordinates. We have run the void identification and stacking algorithm on the particles of this box. Additionally, we have considered the stacked particles when peculiar velocities are either included (dubbed ``mock catalog with full redshifts'') or excluded (dubbed ``mock catalog with pure expansion redshifts''). 

In Figure~\ref{fig:redshift_vs_pure_expansion}, we show the result of the stacking for one of the simulations, for the snapshot at redshift $z=1$. For visualization purposes, the effects of expansions were removed  \emph{after} having stacked the voids. In the left panel, we show the result of the stacking algorithm for the mock catalog with pure expansion redshifts. In the right panel, we show the same test but for mock catalogs with full redshifts. We highlight with black contours the fitted three-dimensional density profile of Section~\ref{sec:stack_comoving}. The  plot is corrected for the inferred expansion, so the contours are perfectly circular. Qualitatively, they match the color coded density in the left panel. 

We note in the right panel of Figure~\ref{fig:redshift_vs_pure_expansion} that there is a non-trivial deformation of the void. At low $(d,z)$, the void is emptier and slightly elongated in the redshift direction. This is expected because of a finger-of-god effect in the void. At high $(d,z)$, the void is flattened.  This is clearly seen by considering, {\it e.g.}, the yellow iso-density in the right panel. The decrease of the peculiar velocities are not sufficient to explain the amplitude of this effect. We have tested our fitting procedure on mock voids whose shaped have been transformed by the average peculiar velocity field both expected and measured in voids.  We have found that they are not introducing a significant pancaking effect. 

We have investigated the origin of this systematic effect, which happens to be non-trivial. It may be explained by a two-step process. First, large-scale redshift space distortions induce a selection bias on cosmic voids: cosmic voids with collapsed structures along the angular coordinate direction are slightly disfavoured when they are binned by void sizes. This discrimination is generated by large scale flows that induces a small modification of the void size, which is sufficient to displace voids from one void size bin to another. However, because voids do not have a flat size distribution, it is statistically much less likely to displace a large void into the bin than to displace it out to a bin corresponding to larger voids. The stretching is far less present when large clusters are present along the line of sight, which means that these voids stays in the same void size bin.  This selection effect makes the distribution of large-scale structure outside the void slightly anisotropic. Additionally, as halos are preferentially located at small angular distance and high redshift distance from the void center, the finger-of-gods that they produce cause a thickening of the void wall in the redshift direction, which in turns cause the pancaking of the voids.

At smaller radii, like $\sim$5 \Mpch{}, the stacked void is seemingly spherical because it is not contaminated by any of the two above effects. By considering the void shape near the half-radius from the center therefore minimizes possible biases due to peculiar velocities. This is a robust procedure, but it is not lossless. Further modeling of the profile could improve signal to noise in our shape inference, described in the next section.  In Section~\ref{sec:hubble_diagram}, we propose a simple alternative de-biasing scheme.

\subsection{Void shape inference in redshift space}
\label{sec:ellipsoidal}

Stacked voids have an isotropic cubic density profile in comoving coordinates, as found in Section~\ref{sec:stack_comoving}.  We model the redshift space distortion by fitting the density $n(d,z)$ estimated using the stacking procedure of Section~\ref{sec:void_stacking}, in redshift/angular coordinates this time, to the function
\begin{equation}
  n(d,z) = \text{min}\left(n_0 + \left((d/a_d)^2 + (z/a_z)^2\right)^{3/2}, n_\text{max}\right) \label{eq:stretched_cubic}
\end{equation}
with $n_0$ the density at the minimum in $(d,z)=(0,0)$, $a_d$ the semi-axis along the angular coordinate direction, $a_z$ the semi-axis the redshift direction, $n_\text{max}$ a maximum density value. We expect $n_\text{max}$ ought to be near unity to show convergence to the mean density. However, we leave it as a free parameter because of the presence of the shell around the void and the limited accessible volume around the stacked void which could bias this value. 

To fit the model to the estimated density we assume that the fluctuations according to the ellipsoidal model are Gaussian but with two different variances depending on the location in the void. The likelihood $\chi^2$ takes thus the following shape:
\begin{multline}
  \chi^2(n_0, n_\mathrm{max}, a_d, a_z, \sigma_0, \sigma_1) = \\ 
    \sum_{i=1}^{N_d} \sum_{j=1}^{N_z} S_{i,j} \left(
    \frac{\left(n(d_i,z_j)-n_{i,j}\right)^2}{\sigma^2(d_i,z_i)} + 2\log(\sigma(d_i,z_i)) \right), \label{eq:chi2_shape}
\end{multline}
with $(i,j)$ the $i$-th and $j$-th bin, for which the $(d,z)$ take the value ($d_i$, $z_j$), $n_{i,j}$ the value estimated in the bin $(i,j)$, $N_d$ the number of bins in the $d$ direction, $N_z$ the number of bin in the $z$ direction. $S_{i,j}$ is either one or zero depending whether we want to include the bin $(i,j)$ in the optimization. We limit the fit to the disc of radius $R_\mathrm{cut}$, as there is no stacked void data at distance larger than this. $S_{i,j}$ takes the form
\begin{equation}
	S_{i,j} = \left\{ \begin{array}{ll}
				1  & \mathrm{if }\;\sqrt{d_i^2 + (z_j/E)^2} \le R_\mathrm{cut} \\
				0 & \mathrm{otherwise}
			\end{array} \right.,
\end{equation}
where we may correct for the expansion in the pixel selection through the coefficient $E$. In practice, we keep $E=1$ in all the following. 

At the position $(d,z)$, and within the void, we enforce that the Gaussian part, $\sigma$, of the distribution $\mathcal{R}$ scales as $1/\sqrt{d}$. This follows from the cylindrical averaging when building the stack of voids. Outside the void, we take a fixed standard deviation to account for uncertainty in approximating the outside profile by a single constant. The standard deviation is spatially varying according to:
\begin{equation}
  \sigma(d,z) = \left\{ 
      \begin{array}{ll}
        \sigma_0 \sqrt{\frac{1 h^{-1}\,\mathrm{Mpc}}{d}} & \mathrm{if}\, n(d,z) < n_\mathrm{max} \\
        \sigma_1 & \mathrm{otherwise}
      \end{array}
      \right.
\end{equation}
 We find the  
parameters $(n_0, n_\text{max}, a_d, a_z, \sigma_0, \sigma_1)$ and their error bars by running a Monte-Carlo Markov-Chain exploration of the four parameters on the sub-sample of pixels for which $\sqrt{d_i^2 + (z_j/E)^2} \le R_\mathrm{cut}$. 
All the measurements of ratios shown in the Section~\ref{sec:hubble} are made using this technique.

The error model contains two components---the Poisson error due to the number of tracers in each pixel, and a correlated error due to dense clumps which occurs around individual voids and which have not been completely removed by the averaging procedure. Such clumps of scale $\lesssim 1$\Mpch{} are visible in Figure~\ref{fig:redshift_vs_pure_expansion}. A crude way to model such correlated fluctuations is to choose the pixel size to be large enough that such fluctuations mostly affect single pixels only.  We conservatively choose a pixel size of 2\Mpch{} to satisfy this criterion. For pixels of this size the Poisson error due to individual tracers is entirely negligible.

\begin{figure}
  \includegraphics[width=\hsize]{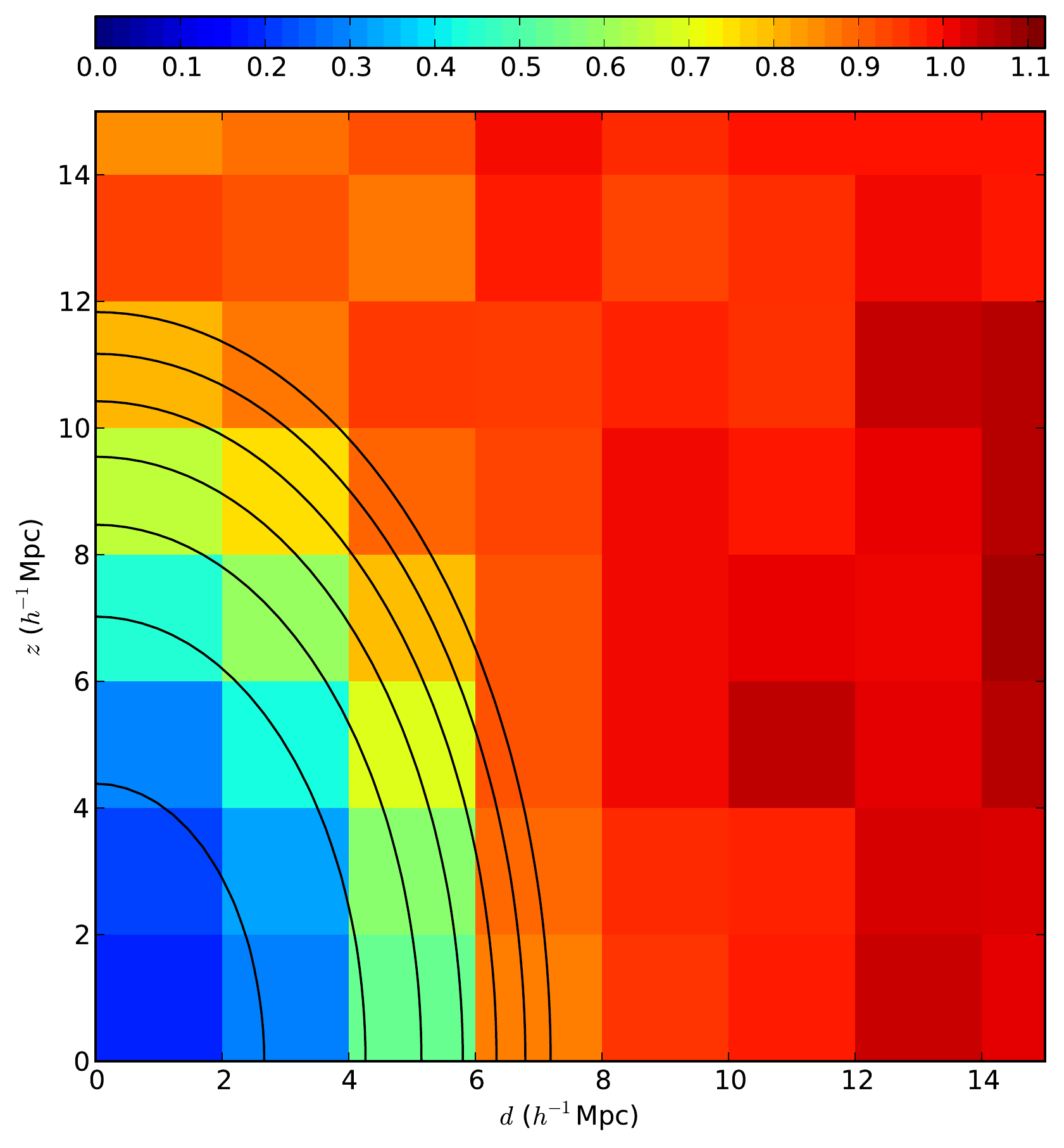}
  \caption{\label{fig:likelihood_stretch} {\it Result of the likelihood analysis on the binned particles of the stacked void} -- We give in solid black line the iso-density contour of the stretched cubic model (Eq.~\ref{eq:stretched_cubic}) fit using the likelihood analysis of Section~\ref{sec:ellipsoidal}. For this Figure, we have considered voids with $r_\mathrm{eff}=8$\Mpch{}. We plot the underlying density field between the null density and 1.1$\bar{\rho}$, with $\bar{\rho}$ the mean density of the slice. Density pixels are 2\Mpch{}. The error is modeled as independent from pixel to pixel. }
\end{figure}

We illustrate in Figure~\ref{fig:likelihood_stretch} the outcome of the likelihood analysis on a stretched stacked void, obtained from a mock light-cone at $z=1$. We show both the density field of the stacked void and the iso-density contours of the fitted profile using the likelihood analysis.  
The iso-density contours follow the outer edge of the void. The fluctuations of the binned density field looks clearly random and uncorrelated at this resolution. This was not the case in Figure~\ref{fig:redshift_vs_pure_expansion}.  We conclude that the likelihood thus behaves as designed. 

We tested the robustness of this procedure to changes in the number density of tracers by re-running the entire pipeline on    a subsample of the $N$-body particles. The results essentially did not change  if the number of particles was reduced by a factor 5 (to $\sim 0.2 h^{3}\text{Mpc}^{-3}$), which corresponds to a typical galaxy density expected for the EUCLID survey at low redshift ($z \la 0.1$). This confirms that Poisson error due to the number of tracers is negligible for the large pixel sizes required to  reduce pixel-to-pixel correlations due to non-linear structures.

\subsection{Number of voids}

\begin{figure}
  \includegraphics[width=\hsize]{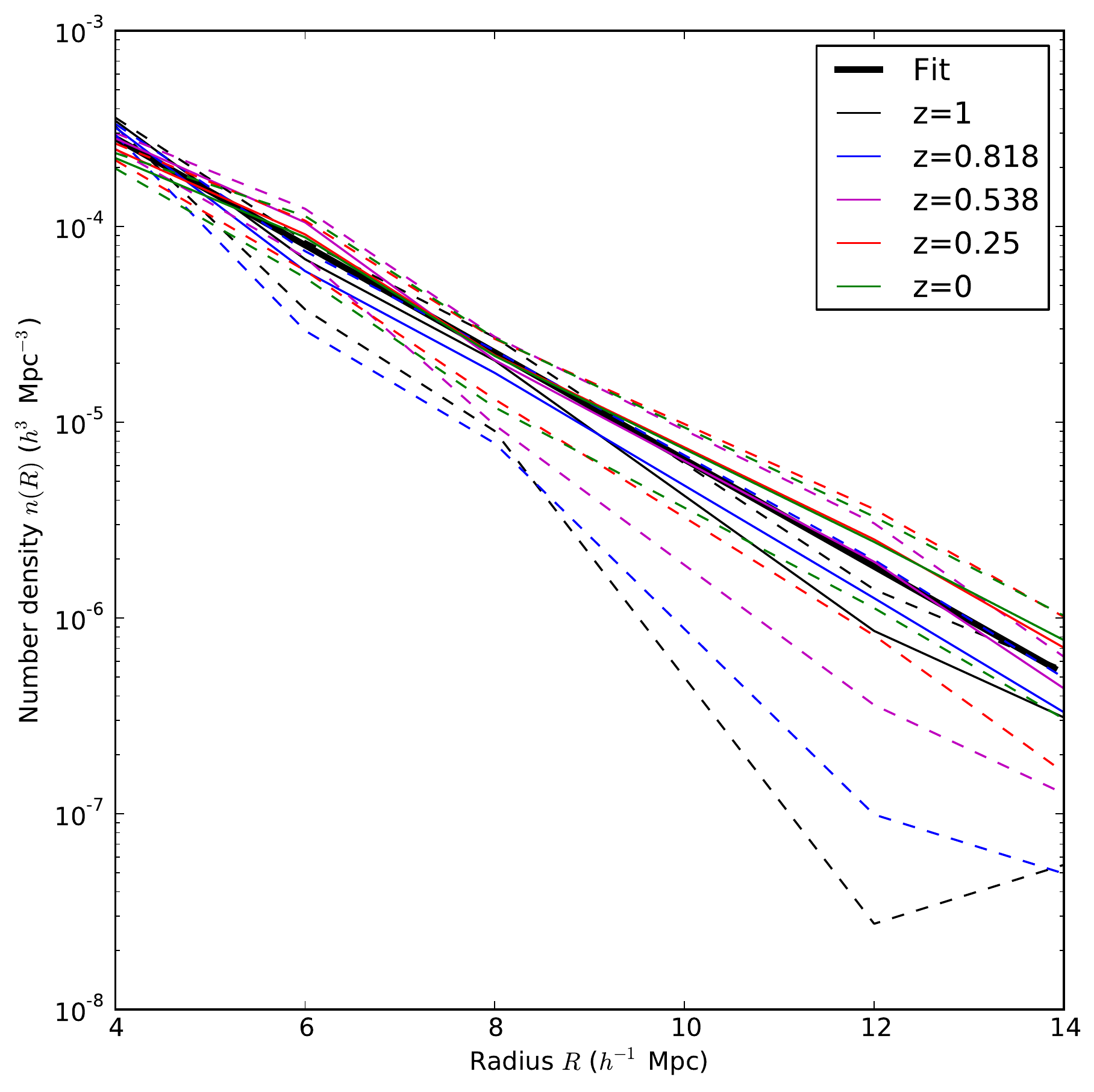}
  \caption{\label{fig:void_number} \textit{Comoving number density of voids.} We show results for purely expanding universes, the result when peculiar velocities contaminates redshifts is the same. The dashed lines give the lowest and the highest value of the density when the three simulations are considered.}
\end{figure}

We give in Figure~\ref{fig:void_number} the averaged number density of voids, in comoving coordinates, for the three simulations, at each redshift and each $r_\mathrm{eff}$ that we have considered for estimating the Hubble constant.

\begin{figure*}
   \begin{center}
	\includegraphics[width=\hsize]{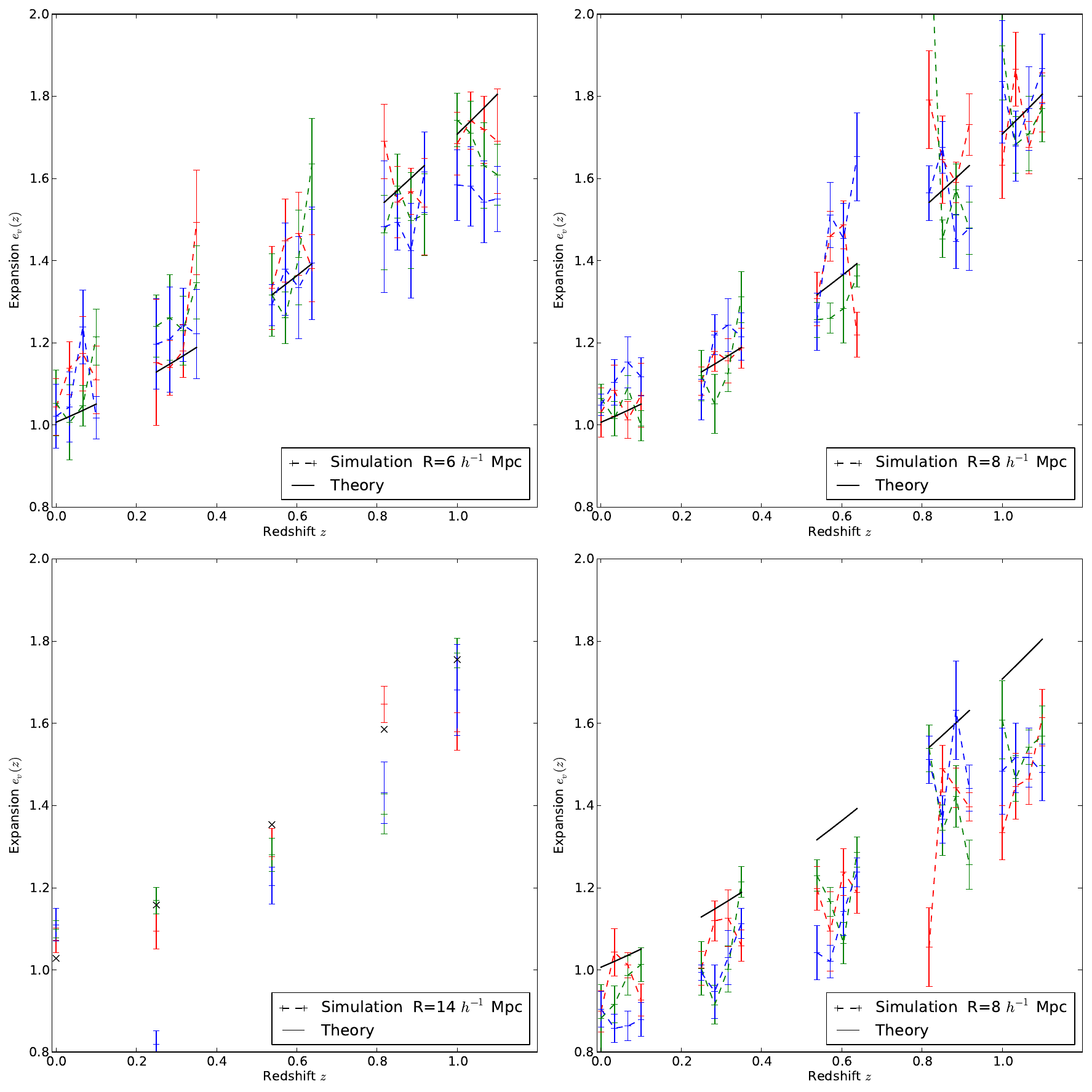}
   \end{center}
	\caption{\label{fig:hubble_diagram} {\it Hubble diagram derived from the voids.} Hubble diagrams derived from voids with three effective radii, without including distortions due to peculiar velocities: 6\Mpch{} (top-left), 8\Mpch{} (top-right), 14\Mpch{} (bottom-left). The bottom-right panel corresponds to $r_\mathrm{eff}=8$\Mpch{} and including distortions due to peculiar velocities. We show the actual expansion in the mock catalogs (black line) and the recovered average expansion from stacked void shapes (colored dashed lines) for the three $N$-body simulations. The colored error bars show the standard deviation inferred using our statistical model described in Eq.~\ref{eq:chi2_shape}. In the bottom right panel, we used a 40,000\kms{} thick slice instead of 10,000\kms{}. }
\end{figure*}

The dependence of the number of voids with redshift is as expected. Small voids should be more abundant at high redshifts and large voids more abundant at low redshifts \citep{SW04}. The relation pivots about a radius of $\sim$8\Mpch{}. As noted by these same authors, the number density of voids depends principally on their volume and then on redshifts through growth of structures. The number of voids detected, when peculiar velocities contamination is added in our mock catalogs, is roughly the same and with the same dependence with radii. We note a small but systematic destruction of voids of 4\Mpch{}. It is plausible that the original topology is lost at these small scales because of the contamination by fingers of god, which can be effectively as deep as 10\Mpch{}. 

In Figure~\ref{fig:void_number}, we note that whatever the dependence of the physical number density of voids with redshifts, for voids \emph{as we define them in redshift space}, this dependence is small within the redshift range $z=0-1$.  Thus for the purpose of this work, we neglect the time dependence of the void abundances and focus on the scale dependence. For each radius, we average the densities at all redshifts and fit an exponential law, which seemed most suited for representing this set of curves, on the number density as a function of $r_\mathrm{eff}$, for the range $4-14$\Mpch{}:
\begin{multline}
	\frac{n(r_\mathrm{eff})}{1\;h^3\,\text{Mpc}^{-3}} = \\ (3.5 \pm 0.2) 10^{-3}\; \exp\left(-(0.632 \pm 0.006) \frac{r_\mathrm{eff}}{1 h^{-1}\mathrm{Mpc}}\right).
\end{multline}
We show this relation in Figure~\ref{fig:void_number} with a thick black solid line.
This relation is used for the Fisher-Matrix analysis of the Section~\ref{sec:fisher_matrix} as an approximation of the behavior of the number of voids as a function of scale. We establish this empirical relation without fitting to any void formation models, such as in the one in \cite{SW04}. This relation only reflects how voids are defined by our algorithm described in Section~\ref{sec:finding_voids}.

\section{Estimating the expansion history using voids}
\label{sec:hubble}

With what precision can we obtain $e_v(z)$, which is closely related to $H(z)$, by only considering the  redshift shape deformation of stacked voids? We use the algorithms and results of Section~\ref{sec:finding_voids} and \ref{sec:test_nbody}. In Section~\ref{sec:hubble_diagram}, we present and discuss the quality of the derived Hubble diagrams. In Section~\ref{sec:fisher_matrix}, we construct a Fisher-matrix analysis of the constraints that can be obtained on Dark Energy with this method, and compare it to expected baryonic acoustic oscillations constraints from the Dark-Energy Task Force \citep{detf}.

\subsection{Results from simulations}
\label{sec:hubble_diagram}

We show in Figure~\ref{fig:hubble_diagram} the complete ``Hubble'' diagram obtained from voids of radius 6\Mpch{} (top-left panel), 8\Mpch{} (top-right panel) and 14\Mpch{} (bottom-left panel). These voids were obtained from the simulations presented in Section~\ref{sec:test_nbody} using the algorithm described in Section~\ref{sec:finding_voids}. For all these diagrams, we have not contaminated the cosmological redshifts by peculiar velocities. Thus they simulate the measurements in a purely expanding universe. We show in the bottom right panel the Hubble diagram for voids of 8\Mpch{}, when peculiar velocities are included in redshifts. We show the actual measurements on the three $N$-body samples, alongside with the 68\% error bar as derived from the Bayesian redshift shape inference. Each color corresponds to the same simulation in all panels. 

\begin{figure}
  \includegraphics[width=\hsize]{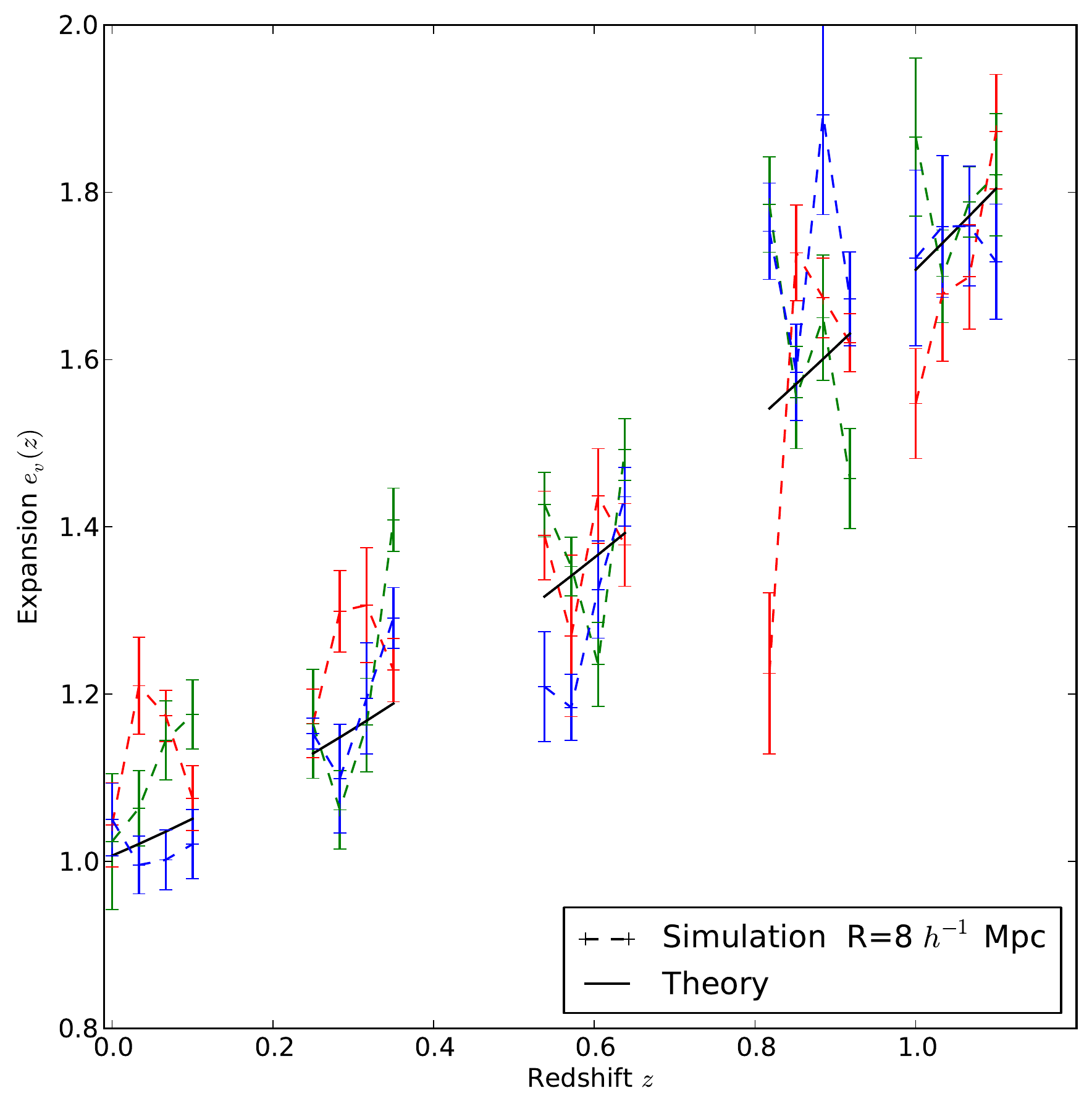}
  \caption{\label{fig:hubblediag_corrected} {\it De-biased Hubble diagram } -- Same as Figure~\ref{fig:hubble_diagram}. We show the Hubble diagram derived from voids with $r_\mathrm{eff}=8$\Mpch{}, including peculiar velocities distortions and after correction of the systematic bias. }
\end{figure}

We have used five snapshots of the simulation, corresponding to time, expressed in redshift, $z=0$, $z=0.25$, $z=0.538$, $z=0.818$ and $z=1$. For each of these snapshots, we have applied cosmological expansion as explained in Section~\ref{sec:stack_all}. For voids of 6\Mpch{} and 8\Mpch{}, we divide the volume in four slices of 10,000~\kms{} according to the redshift direction. For voids of 14\Mpch{}, we keep the full volume of 400\Mpch{}$\times$400\Mpch{}$\times$40,000~\kms{}. 

In black solid line, we plot the average local stretching $\bar{E}(z,\Delta z)$, derived from Eq.~\eqref{eq:local_hubble_comoving},
\begin{equation}
	\frac{1}{\bar{E}(z)} = \left\langle \frac{\delta r}{\delta z} \right\rangle = \frac{1}{\Delta z} \int_{z}^{z+\Delta z} \frac{\mathrm{d}\tilde{z}}{E(\tilde{z})},
\end{equation}
where $z$ is the minimum redshift of the slice and $\Delta z$ its thickness. Assuming that voids are uniformly distributed in the volume, the stacked voids should be stretched by $\bar{E}$. The theoretical expectation and the measurement in $N$-body simulation are in agreement for the three void sizes 6\Mpch{}, 8\Mpch{} and 14\Mpch{}, as clearly shown by the comparison in Figure~\ref{fig:hubble_diagram} (black solid line). We note that the standard deviation derived from the posterior are quite larger than the residual for the voids with $r_\mathrm{eff}=6$\Mpch{}. This is linked to our binning choice and the need for a better model of variations of the density at scales smaller than 2\Mpch{}.

Considering the Hubble diagram obtained when the distortion due to peculiar velocities are included, we note an important systematic effect in the bottom right panel of Figure~\ref{fig:hubble_diagram}. This distortion is a consequence of the pancaking effect that has been discussed in Section~\ref{sec:stack_all}. It is unfortunately non-trivial to model.

The pancaking effect does not affect the overall stretching but it transforms the structure of the density field outside the void, which may bias our statistical estimator of the shape.  It is also expected to be  weakly dependent on cosmology, essentially $\sigma_8$ and $\Omega_\text{m}$ which both affects the amplitude of peculiar velocities inside and outside clusters. We have estimated using mock catalogs from which finger-of-gods were removed that it corresponds to a $\sim$5\% effect on the stretching. The thickening of the walls by the finger-of-god corresponds to the rest of the apparent flattening ($\sim$11\%). On one hand, this systematic should depend slightly on voids sizes as larger voids have a comparatively smaller flattening induced by the finger-of-god. On the other hand, because of the much smaller number of voids, the error bars looks sufficiently big to hide this effect in the present study. We propose to address this issue in a future work focused on mock galaxy catalogs.

As a first approximation we take the bias induced by the pancaking as constant, and leave more detailed statistical and dynamical modeling  for future work. We empirically find that the observed stretching $E(z)$ ($e_v$ in observations) should be multiplied by the constant de-biasing factor of $1.16\pm 0.04$. We have checked that this constant is not strongly dependent on void radii. We note that this multiplicative factor is unimportant for the actual determination of the equation of state of Dark Energy. This statement is correct as long as the pancaking effect is not redshift dependent, which could alter the slope of the relative between redshift and the stretching. Consequently, though we adopt a factor of $1.16 \pm 0.04$ for this section, this factor should be left as a free parameter in any attempt to fit the observations of void ellipticities.

We show the resulting diagram in Figure~\ref{fig:hubblediag_corrected}. We note that the effect has now disappeared except at the lowest redshift where this leads to a slight overstretching. This overstretching is due to a competing effect only present at small redshift and for small voids. 

\subsection{Fisher-matrix analysis for Dark Energy}
\label{sec:fisher_matrix}

To assess the power of the \APz test using stacked voids, we derive the Fisher matrix for the Dark energy properties that can be derived from the Hubble constant. We consider a Chevalier-Polarski-Linder (CPL) parametrization \citep{CP01,Linder03} of Dark energy equation of state:
\begin{equation}
  w(z) = w_0 + w_a \frac{z}{1+z}\label{eq:CPL}
\end{equation}
For sufficiently low redshift the reduced Hubble constant is thus
\begin{multline}
  E(z,w_0,w_a) = \\  \left( \frac{\omega_\text{m}}{h_0^2} (1+z)^3 + \Omega_\Lambda \exp\left(-3\int_{0}^{z}\text{d}z'\,\frac{1+w(z')}{1+z'} \right)\right)^{1/2},
\end{multline}
with $w(z)$ as defined in Eq.~\eqref{eq:CPL}, and $\omega_\text{m}=\Omega_\text{m} h^2$. We do not assume that the cosmology is flat.

\begin{table*}
  \begin{center}
    \begin{tabular}{cccccc}
       \hline
       Survey & Fraction  & Luminosity  & Limiting  & $z_\text{max}$ & Number \\
              & of sky    & function & magnitude &  & of galaxies \\
       \hline
       \multirow{4}{*}{SDSS-DR7} & \multirow{4}{*}{24\%} & $\phi_* = 1.46\,10^{-2}\; h^3$Mpc$^{-3}$ & \multirow{4}{*}{$r=17.77$} & \multirow{4}{*}{$0.3$} & \multirow{4}{*}{$1.7\,10^6$}\\
       & & $M_*=-20.83$  \\
       & & $\alpha=-1.20$ \\
       & & \citep{Blanton00} \\
       \hline
       \multirow{4}{*}{SDSS-DR7 (LRG)} & \multirow{4}{*}{24\%} & $\phi_* = 2.63\,10^{-5}\;h^3$Mpc$^{-3}$ & \multirow{4}{*}{$r=19.5$} & \multirow{4}{*}{$0.45$} & \multirow{4}{*}{$10^5$} \\
       & & $M_* = -19.42$ \\
       & & $\alpha=3.90$ \\
       & & \citep{CEFF08} \\
       \hline
       BOSS & 24\% & same as the SDSS & $r=20$ & $0.7$ & $1.5\;10^6$\\
       \hline
       \multirow{4}{*}{EUCLID} & \multirow{4}{*}{36\%} & $\phi_* = 1.16\,10^{-2}\;h^3$Mpc$^{-3}$ & \multirow{4}{*}{$H=24$} & \multirow{4}{*}{$1.5$} & \multirow{4}{*}{$\sim 1.6\, 10^8$} \\
       & & $M_* = -23.39$ \\
       & & $\alpha=-1.09$ \\
       & & \citep{Kochanek2001,JPCS06} \\
       \hline
    \end{tabular}
  \end{center}
  \caption{\label{tab:surveys} Survey parameters}
\end{table*}

We consider a hypothetical measurement of the stretching constant $e_\mathrm{v}(z)$ from the study of voids.  Thus we have an estimate of $e_{\mathrm{v}}(z)$ at different redshifts $z_i$ that we label $e_{\mathrm{v},i}$. Each of the estimates has an independent random error variance $V_i$. The likelihood $\mathcal{L}$ of the cosmological parameters $p$ is thus simply described by:
\begin{equation}
  \mathcal{L}(p) = \sum_{i=1}^{N_z} \frac{\left(e_{\mathrm{v},i} - \mathrm{e}_{\mathrm{v}}(z_i, p)\right)^2}{V_i}.
\end{equation}
In $p$, we include a parameter, $b_v$, which corresponds to the overall flattening of the voids due to effects of redshift space distortions. We use the value and the uncertainty as determined in Section~\ref{sec:hubble}. This prior is of no consequence on the Fisher-Matrix derived for the EUCLID survey as it is sufficiently dense to constrain its value at lower redshift. The definition of the Fisher matrix is
\begin{equation}
  F_{k,l} = \left\langle \frac{\partial \mathcal{L}}{\partial p_k} \frac{\partial \mathcal{L}}{\partial p_l} \right\rangle,
\end{equation}
with $k,l \in p$ and the averaging is taken over all the possible realizations of the data-sets $\{ e_{\mathrm{v},i} \}$ given the cosmological parameters $p$ and the noise determined by $\{ V_i \}$.
In our case, this definition simplifies to:
\begin{equation}
	F_{k,l} = \sum_{i=1}^{N_z} \frac{1}{V_i} \frac{\partial e_\mathrm{v}}{\partial p_k} \frac{\partial e_\mathrm{v}}{\partial p_l}.
\end{equation}

To complete the evaluation it is required to have an estimate of $V_i=\langle (e_{\mathrm{v},i}-e_\mathrm{v}(z_i))^2\rangle$, the variance of the estimated Hubble constant in each redshift slice. This variance may have the following typical dependence:
\begin{equation}
	V_i = \left(\sum_{j} \left(\epsilon(R_j, N(R_j,z))\right)^2\right)^{-1},\label{eq:mock_variance}
\end{equation}
with $\epsilon(R,N)$ the standard deviation of the estimator if we stack $N$ voids with a size $R$, $\bar{n}(R,z)$ the number density of voids of size $R$ at redshift $z$. In Eq.~\eqref{eq:mock_variance}, we are summing over all the void effective radii that are observable in the slice $i$. If we bin void by size, they may form statistically independent stacks. It is thus possible to improve the variance of the local expansion factor $\tilde{E}(z)$ by considering all possible sizes at once. From the tests that we have run in Section~\ref{sec:test_nbody}, we know that binning voids in bins of $\Delta=1$\Mpch{} in effective radii gives adequate results. Thus, in Eq.~\eqref{eq:mock_variance}, we use
\begin{equation}
  R_j = R_\text{min obs.}(z) + j \Delta,\label{eq:void_size_list}
\end{equation}
with $j < (R_\text{max obs.}-R_\text{min obs.})/\Delta$.  Of course, it is not possible to observe void which have a size larger than half the width of the slice $i$. Consequently,
\begin{equation}
  R_\text{max obs.} = \frac{c \delta z}{2 H_0},
\end{equation}
with $\delta z$ the width of the slice, expressed in redshift units. We note that this is only an approximation. 

We have chosen the following scaling for the minimal observable void size $R_\text{min obs.}$ used in Eq.~\eqref{eq:void_size_list}
\begin{equation}
	R_\text{min obs.}(z) = \text{min}\left( 6 h^{-1}\text{ Mpc}; s_r n(z_i)^{-1/3} \right), \label{eq:min_void_size}
\end{equation}
with $n(z)$ the mean comoving density of galaxies at redshift $z$. $n(z)$ typically depends on $\phi(M)$, the luminosity function of galaxies in the observed band. For a magnitude limited catalog, as the SDSS Main galaxy sample, assuming magnitudes are corrected for evolution $n(z)$ is related to $\phi(M)$ as
\begin{equation}
   n(z) = \int_{-\infty}^{m - 5\log_{10}(d(z))} \phi(M)\;\mathrm{d}M \label{eq:lf_numden}
\end{equation}
Our choice of $R_\text{min obs.}$ ensures that the size of the structures that are observed is limited by the number density of tracers. The Fisher matrix, and of course the figure of merit for the determination of Dark Energy, is going to depend on $s_r$. We discuss in Section~\ref{sec:fisher_results}, the impact of varying $s_r$. We do not accept voids smaller than 6\Mpch{} because they may be easily disrupted and strongly contaminated, in redshift coordinates,  by distortions due to peculiar velocities and are more difficult to identify. We leave the exact determination of the lower limit of observable voids for future work.

We approximate the number of voids in the slice of thickness $\delta z$ at redshift $z$ by
\begin{equation}
	N = \frac{4 \pi}{3} f_\text{sky} \left(\frac{c}{H_0}\right)^3 \left(\chi(z+\delta z)^3-\chi(z)^3 \right) \bar{n}_\text{void}(R,z).
\end{equation}
We determine $\epsilon$, $\bar{n}$ from the three simulations that we have run for a $\Lambda$CDM-WMAP7 cosmology. As this cosmological model gives a good fit to CMB and large-scale structures observations \citep{WMAP7_K11,LRG_DR7}, it should give a fair representation of the statistical structure of the galaxy redshift survey.

\begin{figure}
    \includegraphics[width=\hsize]{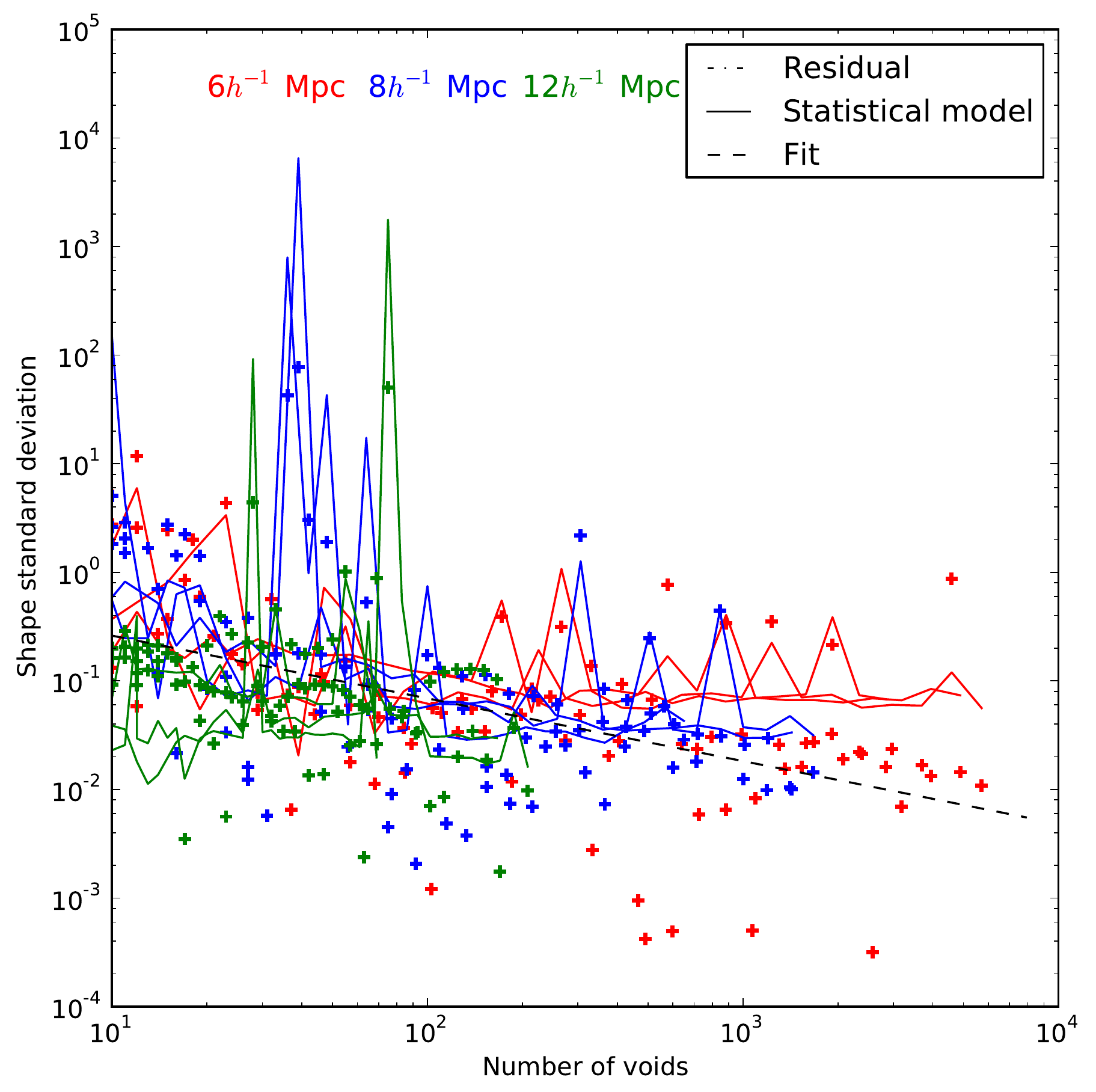}
	\caption{\label{fig:epsilon_variance} {\it Standard deviation of void apparent shape vs the number of stacked voids.} Shape errors inferred by our statistical model  Eq.~\eqref{eq:chi2_shape} (solid lines), the residual between the expected shape and the actual best estimate (thick cross markers). We also show with a dashed line the best fit of a power-law on the residual (Eq.~\ref{eq:variance_reduction_ref}). The results for the three $N$-body simulations are shown here. The colors correspond  to voids with an effective radii of 6\Mpch{} (red), 8\Mpch{} (blue) and 12\Mpch{} (green). }
\end{figure}

For the three void sizes that we have considered in this work, $6$\Mpch{}, $8$\Mpch, and $14$\Mpch, we show in Figure~\ref{fig:epsilon_variance} the function $\epsilon(R,N)$. We may approximate this function by:
\begin{equation}
	\epsilon(R,N) \simeq \left(\frac{0.98}{N}\right)^{-0.57} \label{eq:variance_reduction_ref},
\end{equation}
which is independent of the radius $R$.
The reduction in the variance is scaling roughly as a Poisson distribution. This corresponds to the expectation that the noise in the shape estimation comes essentially from the void-to-void fluctuations. For future surveys we expect to observe a huge number of small voids which would make our estimate sensitive to the low error cases, corresponding to the red points Figure~\ref{fig:epsilon_variance}. Conservatively, we thus take that the standard deviation is scaling exactly as
\begin{equation}
  \epsilon(N) = \frac{1}{\sqrt{N}}\label{eq:variance_reduction}.
\end{equation}
This  slightly overestimates the errors for the shape measurement and should give  a conservative estimate of the expected constraints.

\subsection{Application to present and future surveys}
\label{sec:surveys}
\label{sec:fisher_results}

\begin{figure}[t]
	\includegraphics[width=.9\hsize]{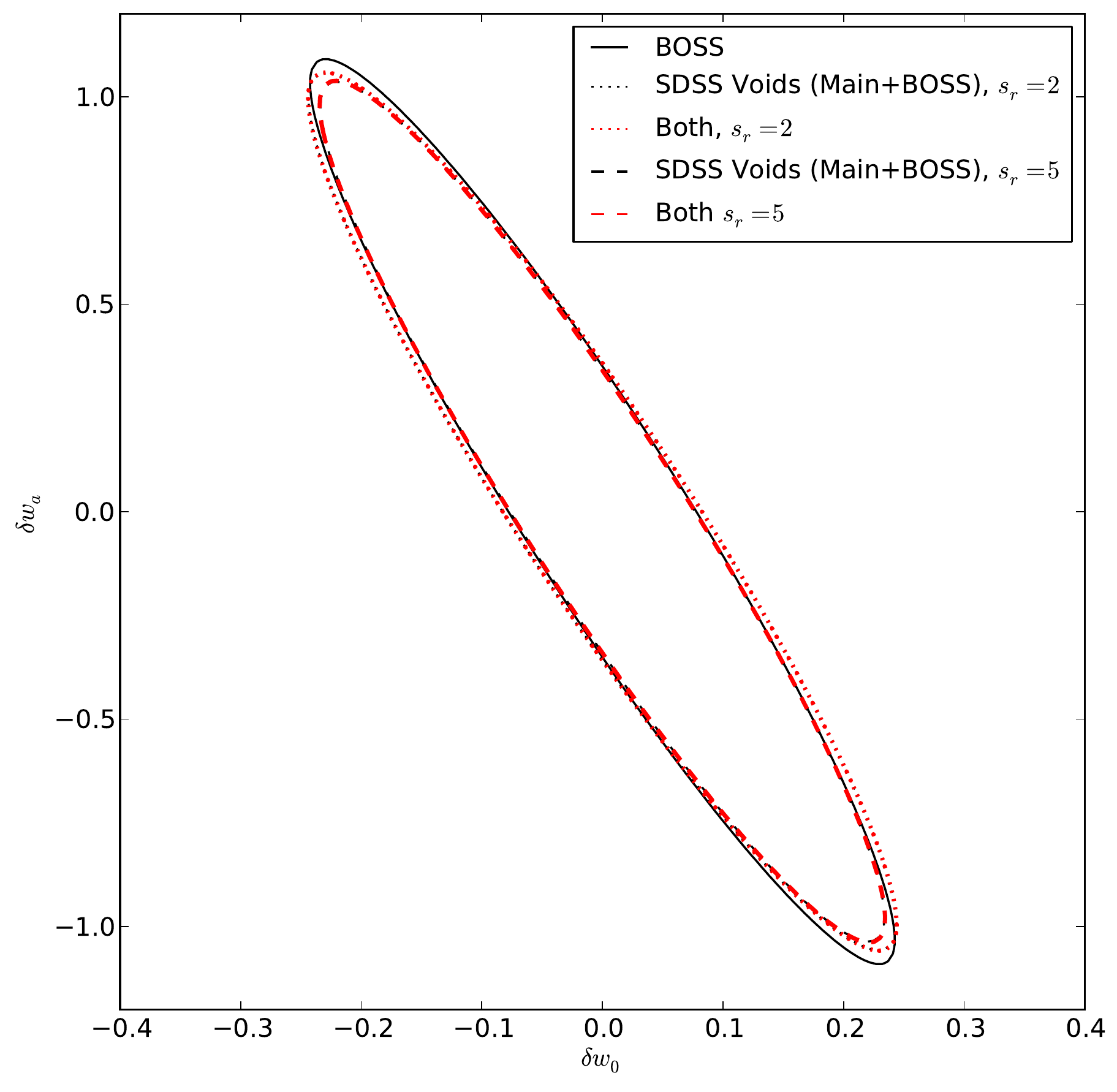}
	\caption{\label{fig:fisher_constraints_boss} {\it Fisher matrix forecasts for SDSS/BOSS survey.}  Predicted 95\% confidence regions, assuming {\sc Planck} prior and the determination of the Hubble constant $H=72 \pm 8$~\kmsMpc. The {\sc Planck} priors were obtained using the procedure in the report of the DETF. Please note that we plot $w_0$ vs $w_a$ and not $w_p$ vs $w_a$.  The black solid line gives the constraints derived from the analysis of baryonic acoustic oscillations in the BOSS survey. The dotted lines gives the constraints derived from the shapes of stacked voids using SDSS main galaxy sample and the BOSS sample. The dashed line shows the obtained constraint by combining both Baryonic Acoustic Oscillations and void shape measurement. The constraints are either derived assuming $s_r=2$ (black) or $s_r=5$ (red).}
\end{figure}

In this section, we move on to apply the formalism of Section~\ref{sec:fisher_matrix} to three surveys: the SDSS-DR7 \citep{SDSS_DR7}, the BOSS survey \citep{BOSS} and the EUCLID survey \citep{EUCLID2}. We use slices of $\delta z = 0.03$, which corresponds approximately to a comoving thickness of 90\Mpch{}. This limits the effective radius of the void to $\sim 45$\Mpch{}.

Table~\ref{tab:surveys} lists the parameters of those surveys which are useful for the Fisher-Matrix analysis. These parameters can be translated, using Eq.~\eqref{eq:lf_numden}, in a number density of tracers at a given redshift. This determines the minimal size of observable voids through Eq.~\eqref{eq:min_void_size}. Using Eq.~\eqref{eq:variance_reduction}, we obtain the number density of voids and with the void shape standard-deviation reduction $\epsilon(R,N)$.

For deriving Fisher-matrix constraints, we assume the priors that should come out from the {\sc Planck} mission. We use the Dark Energy Task Force \citep{detf} prescription for deriving these priors. In addition, we apply constraints from Stage II experiments as expressed in the DETF report. The figure of merit (FoM) is defined as 
\begin{equation}
   \mathrm{FoM} = \frac{1}{\sigma(w_a)\sigma(w_p)}
\end{equation}
with
\begin{equation}
   w_p = w_0 + (1-a_p) w_a
\end{equation}
and
\begin{equation}
   1-a_p = -\frac{\langle \delta w_a \delta w_0 \rangle}{\langle (\delta w_a)^2 \rangle}.
\end{equation}
$w_p$ is the best estimate we can obtain on the equation of state of Dark Energy, evaluated at the scale factor $a_p$ of our Universe.

\begin{figure}[t]
	\includegraphics[width=\hsize]{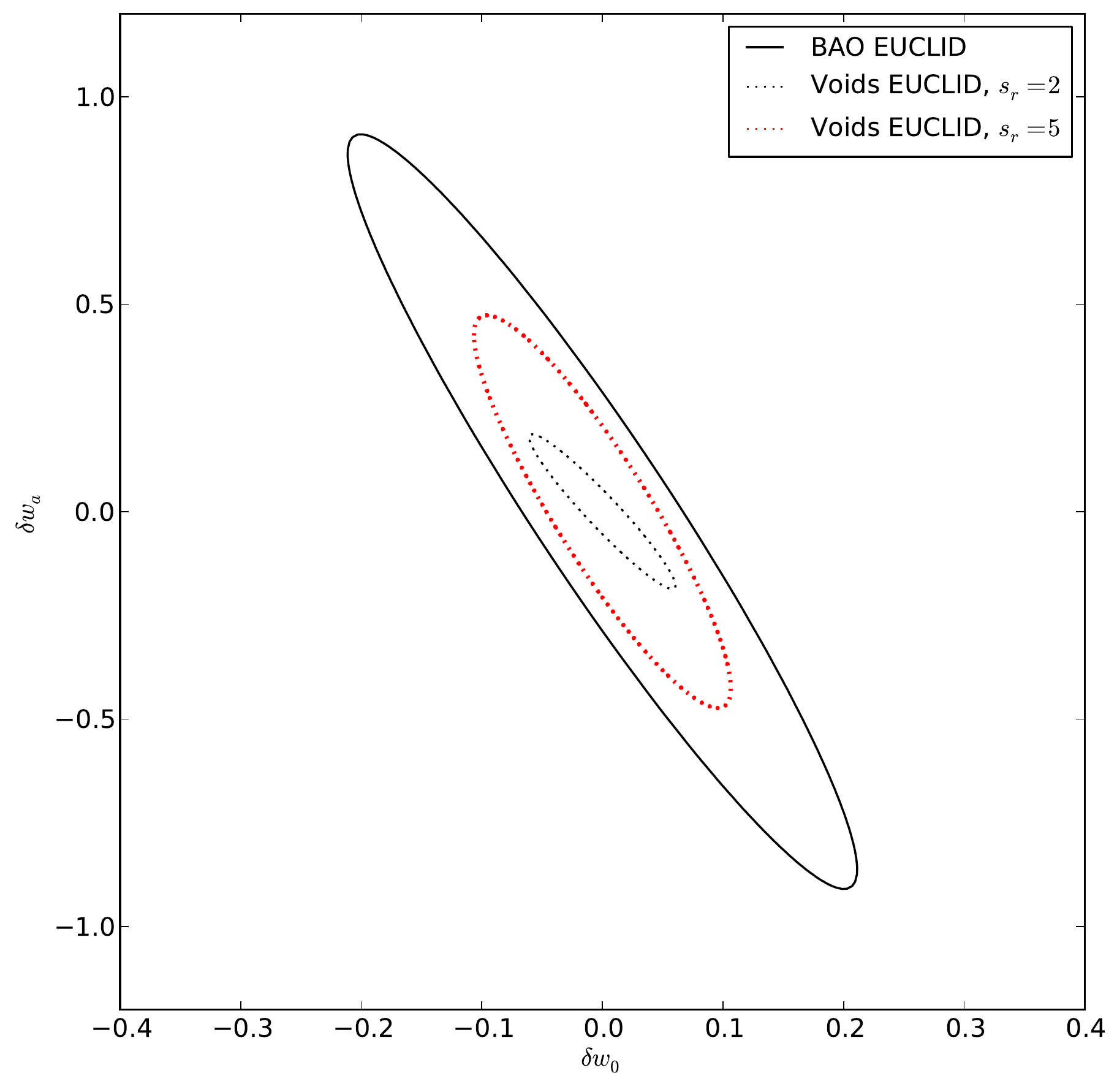}
	\caption{\label{fig:fisher_constraints_euclid}  {\it Fisher matrix forecasts for the EUCLID wide  survey.}  Same as Fig.~\ref{fig:fisher_constraints_boss} but for the EUCLID survey. The solid (dotted) line gives the constraints derived from the analysis of Baryonic Acoustic Oscillations (shapes of stacked voids, assuming $s_r=2$ in black and $s_r=5$ in red).  }
\end{figure}

We consider either an experiment consisting of only BAO analysis, only Voids analysis or the two combined. We do this analysis for the SDSS main galaxy sample, the BOSS survey and the EUCLID survey. We give the results in Table~\ref{tab:fom}, Figure~\ref{fig:fisher_constraints_boss} and Figure~\ref{fig:fisher_constraints_euclid}.

We note that voids and BAO have roughly the same constraining power when considering the SDSS/BOSS experiment (FoM of 71 for BAO only vs. 68 for voids only). The combination of the two slightly improves the constraints (FoM of 75). 
On the other hand, voids are far superior at constraining the equation of state of Dark Energy with the EUCLID survey. We have found a FoM of $\sim$1~825 far superior than the constraints derived from BAO for the same survey. This comes from the possibility of using small scales geometry and not only the $\sim$100\Mpch{} scale which corresponds to BAO.

The results depend on several parameters that we have adopted for the analysis. For example, we have chosen $R_\mathrm{max}=45$\Mpch{}. We have barely found any change in the results by limiting to $r_\mathrm{eff} \simeq 14$\Mpch{} or by changing the thickness $\delta z$ of the slice up to $0.1$. However, changing $R_\text{min}$ have a lot more impact on the FoM. This indicates that, as expected, most of the information comes from the smallest observable voids because of their huge abundance relatively to bigger ones. Similarly, we have tried varying $s_r$ between 1 and 10 to check the stability of the constraints for EUCLID. We have found that the Hubble diagram for voids always give a significant additional information.  As an example, in Figure~\ref{fig:fisher_constraints_euclid}, we show in black (in red respectively) the constraints for $s_r=2$ ($s_r=5$ respectively). The figure of merit may either strongly improve whenever $s_r$ is reduced, up to one hundred times improvement, or diminish by a factor of a few when $s_r$ is increased. We also tried to check the influence of the sky coverage on the FoM. Interestingly, reducing the EUCLID survey to 100~deg$^2$ yields a figure of merit of $\sim$380 for voids, while the one we expect from BAO with BOSS is $\sim$71. Voids could thus yield very good constraints on Dark Energy from deep redshift surveys with smaller sky coverage.

Finally, we note that in all this work we have only used the shape of the \emph{stacked} voids as a probe.  Additionally, there is information in the distribution of void orientations \cite{JF98}.
The distribution of the intrinsic shapes of individual voids contains a great deal of complementary information, at the potential cost of requiring to assume a galaxy bias \citep{PL07,BAW10,LW10}. We think that the intrinsic shape could increase substantially the constraints that only comes from void shape analysis. This could be done in two steps: the first pass would use the method that we have developed in this paper to constrain the local geometry, the second pass would use the shape distribution to constrain the growth of structures.

\begin{table}
  \begin{center}
  \begin{tabular}{ccc}
    \hline
  	Method & Data & FoM  \\
  	\hline
  	BAO & BOSS & 71 \\
  	Voids ($s_r=2$) & SDSS+BOSS LRG & 69 \\
  	Voids ($s_r=5$) &               & 68 \\
  	BAO+Voids ($s_r=2$ or $5$) & SDSS+BOSS & 75 \\
  	Voids ($s_r=2$) & EUCLID &  $\sim$1~825\\
        Voids ($s_r=5$) &        & $\sim$273\\
  	BAO &  & 98 \\
  	BAO+Voids ($s_r=2$) &  & $\sim$2~956 \\
  	BAO+Voids ($s_r=5$) &  & $\sim$380 \\
  	\hline
  \end{tabular}
  \end{center}\noindent
  {\sc Note}: The figure of merit is computed as the non-normalized $1/(\sigma(w_p)\times \sigma(w_a))$ as in the DETF report.  We have included the prior from Stage II dark energy experiments, prior on $H_0$ from the Hubble space telescope and Planck prior.
  \caption{\label{tab:fom} Comparison of figure of merits (FoM)}
 \end{table}

\section{Discussion and Conclusion}
\label{sec:conclusion}

We showed  that by identifying and stacking cosmic voids in redshift shells and size bins, and then measuring their shapes in redshift space we can directly constrain the cosmological expansion through a purely geometric approach. 
Several steps are required to use the stacked voids technique to connect a spectroscopic survey to the expansion geometry and hence to dark energy phenomenology. 
In this paper we proposed methods for each one of these steps which, together, amount to a first  analysis pipeline for the stacked voids technique.

We  use a modified  {\sc Zobov} \citep{N08} algorithm  for finding and stacking voids on a light-cone,  extended   to produce non-overlapping voids  selected according to two criteria: an effective radius within a given range and a central density sufficiently low to mark the region as a void. 

In Section~\ref{sec:test_nbody}, we applied this algorithm to mock light-cone catalogs obtained from three $N$-body simulations. We have tested the method in the original comoving coordinates of the simulation, which has provided us with a model of the  density profile of the stacked void. Then, we have simulated cosmological expansion and distortions due to peculiar velocities, which allowed us to qualitatively estimate the impact on our measurement of the stretching of voids. We find that even a crude de-biasing prescription of the (mild) peculiar velocity systematics yields a powerful method. By generating void stacks for different void sizes and redshift shells, we project out the details of individual void shapes.

We have developed and tested a Gaussian statistical model able to estimate the stretching of the stacked voids. Combining the results obtained from void stacks at all redshift shells results in an estimate of the expansion history.

We are aware that each one of these analysis steps can likely be  improved significantly. There are parameters to optimize, such as the  widths of the redshift shells, and the size bins. We only scratched the surface of possible methods in spatial statistics and computational geometry when it comes to defining voids in realistic surveys. Our method for measuring the redshift space shapes of stacked voids is only one of many that one could imagine. In particular, we have not yet taken advantage of the ability to model the systematics due to peculiar velocities, present at 10-15\% of the expansion signal, except to suggest a crude de-biasing approach which gives a consistent result on our simulations.

Based on these first results we extrapolated and performed Fisher-matrix forecast of the constraints on Dark Energy equation of state we expect from the  SDSS, BOSS and EUCLID spectroscopic surveys.

We have found that cosmic voids have the potential to provide a far more powerful constraint on dark energy than measurements of the Baryonic Acoustic Oscillation scale, by an order of magnitude. This large increase of information is easily understood in comparing the number of modes probed by voids compared to BAOs, which scales roughly as the third power of the ratio of the BAO scale to the scale of the smallest usable voids $\sim 1000$. The area of parameter constraints scales as the square root of the number of modes $\sim 30$. When projected into the $w_{a},w_{p}$ plane using the Fisher matrix formalism for the EUCLID wide survey, we find the improvement over BAO on those parameters by a factor of $\sim 30$. 

We expect our stacked void shape measurements  to be robust to galaxy bias as it is purely geometrical and relies on the topology of the density field \citep{S98}. In fact, it is possible that biased tracers of the density enhance the contrast of voids and therefore enhance the void detection rate.  Another limitation comes from the effective volume accessible from galaxy surveys which is not infinite and may only cover some parts of the sky. If the survey area is sufficiently contiguous with respect to the considered void sizes then the void identification should not be affected by the constraints imposed by the geometry. Prior to the stacking, the identified voids must be rotated such that the stretching direction is always in the same fiducial direction. Finally, we do not expect apparent magnitude limitation to impair the analysis differently than just removing the smallest observable voids at a given redshift. This expectation is based on the conservation of topological properties under resampling of the density field. All these expectations remains to verified on more realistic mock catalogs and real data.

Based on our Fisher matrix forecasts, the stacked voids technique alone promises to double the figure of merit from EUCLID  when compared to the combined results from all other probes using the same data (BAO, weak lensing, type Ia supernovae, cluster counts). The \APz  test using stacked voids is therefore potentially a significant addition to the portfolio of major dark energy probes which merits further detailed  studies focused on additional real-world systematics and optimal survey design.

\section*{Acknowledgments}

 The authors thank Laird Thompson, Joseph Silk, Mark Neyrinck, Thierry Sousbie, Miguel Arag\'on-Calvo, and St\'ephane Colombi  for useful discussions. The authors thank Hans A. Winther for pointing out an inconsistency that we corrected in Figure~3 and 4 in an earlier version of this paper. 

The authors acknowledge financial support from NSF Grant AST 07-08849, AST 09-08693 and from BDW's Chaire d'Excellence granted by the Agence Nationale de Recherche. GL acknowledges support from CITA National Fellowship and financial support from the Government of Canada Post-Doctoral Research Fellowship. This research was supported by the National Science Foundation through TeraGrid resources provided by NCSA under grant number TG-AST100029.

Research at Perimeter Institute is supported by the Government of Canada through Industry Canada and by the Province of Ontario through the Ministry of Research and Innovation.


\end{document}